\def\bq{\begin{equation}}
\def\eq{\end{equation}}
\def\bqa{\begin{eqnarray}}
\def\eqa{\end{eqnarray}}
\def\bqb{\begin{eqnarray*}}
\def\eqb{\end{eqnarray*}}
\def\to{\rightarrow}
\documentstyle[12pt,epsf]{article}
\renewcommand{\theequation}{\arabic{section}.\arabic{equation}}
\global\nulldelimiterspace = 0pt

\def\roughly#1{\mathrel{\raise.3ex
    \hbox{$#1$\kern-.75em\lower1ex\hbox{$\sim$}}}}

\begin{document}
\pagenumbering{arabic}
\thispagestyle{empty}
\def\thefootnote{\fnsymbol{footnote}}
\setcounter{footnote}{1}
 
\begin{flushright}
 CERN-TH/98-172 \\{BI-TP 98/13}
\end{flushright}

\vspace {1 cm}
\begin{center}
{\Large\bf A Simple Born-Form Approximation for $e^+ e^- \to W^+ W^-$ 
at One Loop}
\footnote{This work is supported partly by the Ministry of Education and 
Culture, Japan, under Grant-in-Aid for basic research program C (No.
08649391), by Alexander von Humboldt-Stiftung, Bonn, Germany, 
the BMBF, Bonn, Germany and by the EC contract CHRX-CT94-0579.\\
  {\normalsize
    \begin{flushleft}
       CERN-TH/98-172 \\ June 1998
    \end{flushleft}} 
}
\\
\vspace {1.5 cm}
{\large  M. Kuroda$^a$ and D. Schildknecht$^b$}
\vspace {0.5cm}  \\

$^a$ Institute of Physics, Meiji-Gakuin University\\
Yokohama, Japan.\\
\vspace{0.2cm}
$^b$Theory Division, CERN, Switzerland \\ 
    and\\
Department of Theoretical Physics, University of Bielefeld \\
Bielefeld, Germany\\
\vspace{1cm}
\end{center}
 
\vspace {1cm}

\noindent{\bf Abstract}\par
     A simple Born-form approximation at the one-loop level for 
$e^+e^-\to W^+ W^-$ at 
high energies is given in analytic form.  The different contributions 
to the three invariant one-loop amplitudes, $S_I^{(-)}(s,t)$
and $S_Q^{(\pm)}(s,t)$,
determining the Born-form helicity amplitudes are thoroughly investigated 
analytically and numerically.  At energies  above 500 GeV, the 
accuracy of the simple Born-form 
approximation for the differential production cross section 
is better than $1\%$ for almost all $W^+W^-$ production angles,
independently of whether the $W^+W^-$ polarization is summed over,
or whether a longitudinal or a transverse polarization is
selected for both the $W^+$ and the $W^-$. 
\vspace{0.5cm}

\section{\bf  Introduction}\par
    The process of $e^+e^-$ annihilation into $W$ pairs is one of 
the outstanding reactions to be explored at LEP 2 and at any future 
linear $e^+e^-$ collider.  While simple at tree level, the evaluation 
of the amplitudes for this reaction at the one-loop level \cite{LV}-- 
\cite{BDSBBK}, as a consequence of the large number of contributing
Feynman diagrams, leads to formulae of enormous complication 
only available in extensive computer codes.
A simple approximation of the one-loop result, at least in the 
high-energy limit, is highly desirable from a theoretical point of 
view as well as for practical reasons.\par

     In ref.~\cite{FK3S}, it was conjectured that 
the one-loop helicity amplitudes may be represented, in good approximation,  
by helicity amplitudes that have the form of a Born approximation, only 
differing from the tree-level Born amplitudes by a replacement of the 
electromagnetic and of the weak coupling by $s$- and $t$-dependent invariant 
amplitudes.  The necessary condition of unitarity constraints in the 
high-energy limit acts as a guiding principle for the choice of the set of 
(only) three invariant amplitudes appearing in the Born-form approximation.  
In ref.~\cite{FK3S}, it was indeed shown that  the total cross section and 
the angular distribution of the $W$ bosons may be well approximated by such a 
Born-form approximation. From a slightly different point of view, an 
identical Born-form approximation was given in ref. \cite{DBD}.\par

     The demonstration that a Born-form approximation of the helicity
amplitudes is posssible, in refs. \cite{FK3S},\cite{DBD}, had to rely on a 
numerical evaluation of the one-loop-level amplitudes. No insight 
into the detailed structure of the contributing three invariant 
amplitudes has thus been obtained, and the 
evaluation of the three invariant amplitudes by the corresponding 
computer codes is not much simpler than the evaluation of the 
full one-loop helicity amplitudes, which depend on twelve
invariant amplitudes.  In other words, the important 
problem of deducing a simple (high-energy) approximation for the 
three invariant amplitudes  appearing in the Born-form approximation 
was left unsolved at the time.  Moreover, no attempt was made 
in refs. \cite{FK3S},\cite{DBD} to construct the Born-form approximation 
in such a manner that not only the cross section summed over  
$W$ polarizations, but also the cross
sections for specific $W$ polarizations would be 
adequately approximated.\par
     The purpose of the present work is twofold: \par
     i) to show that 
the three invariant amplitudes appearing in the Born form for 
the helicity amplitudes can indeed be chosen in such a manner that
the helicity amplitudes for fixed polarization of the outgoing $W$-bosons 
are adequately approximated, and \par
     ii) to give a simple analytical 
high-energy approximation for these invariant amplitudes.\par

     A realistic description\footnote{ We refer to ref.~\cite{D1},
\cite{D2}, \cite{BB} and the literature quoted therein.}
 of $W$-pair production requires that 
hard-photon radiation be added and $W$ decay be taken care of 
in conjunction with the inclusion of background processes for which
a treatment at tree level is sufficiently accurate.  In this context 
the present work provides a simple and compact representation 
of the virtual electroweak and soft-photon-radiation corrections to
$W$-pair production.\par
     In section 2, we will give the one-loop Born form which 
at sufficiently high energies yields an adequate approximation of the helicity 
amplitudes for transverse as well as longitudinal polarization of the $W$ 
bosons. In section 3, a brief analytic high-energy approximation 
for the invariant amplitudes $S_I^{(-)}(s,t)$ and $S_Q^{(\pm)}(s,t)$ in the
Born-form approximation will be given.  In section 4, the simple 
high-energy approximation for these amplitudes will be numerically compared 
with the full one-loop results.  In section 5, the exact and the approximated 
differential cross sections for various $W$ polarizations will be compared. Some 
details on the derivation of the Born-form approximation described in
the present paper and a numerical comparison with the previously 
suggested approximation are shifted to appendices A, B and C.  Final conclusions 
will be drawn in section~6.

\medskip
\section{\bf  Born-Form Approximation}\par
     We briefly recall the Born approximation for the process 
$e^+e^- \to W^+W^-$.  The helicity amplitudes may be written 
as a sum of two terms proportional to the squares of the SU(2) 
gauge coupling $g$ and the electromagnetic coupling $e$:
\bq
 {\cal H}(\sigma,\lambda,\bar\lambda) =
   {{g^2}\over 2} {\cal M}_I(\sigma,\lambda,\bar\lambda)\delta_{\sigma,-}
   +e^2 {\cal M}_Q(\sigma,\lambda,\bar\lambda).
\label{2.1}
\eq
Here, and elsewhere, $\sigma$ and $\lambda$, $\bar\lambda$ denote 
twice the electron
helicity and the $W^+$,$W^-$ helicities, respectively.  In the notation of
ref. \cite{FK3S}, the amplitudes ${\cal M}_I$
and ${\cal M}_Q$ are given in terms of the basic amplitudes 
$\bar M_1(\sigma,\lambda,\bar\lambda)$ and 
$\bar M_5(\sigma,\lambda,\bar\lambda)$ by
\bqa
     {\cal M}_I(\sigma,\lambda,\bar\lambda) &=&
      -~{1\over{s-M_Z^2}}\bar M_1(\sigma,\lambda,\bar\lambda)
      -{1\over t}\bar M_5(\sigma,\lambda,\bar\lambda), \label{2.2}\\
     {\cal M}_Q(\sigma,\lambda,\bar\lambda)& =&
 {{M_Z^2}\over{s(s-M_Z^2)}} \bar M_1(\sigma,\lambda,\bar\lambda).
\label{2.3}
\eqa
The evaluation of the cross section requires a choice of the scale at which
$g$ and $e$ are to be related to experiment.  For the high-energy
process of $W$-pair production the choice of a high-energy scale, such as 
the centre-of-mass energy squared $s$, seems most appropriate. 
At LEP2 energies, the choice of $s\cong M_W^2$ was shown to yield
reasonable results \cite{KKS}.  This choice  amounts to employing $e(M_W^2)$
and $g(M_W^2)$, where $g(M_W^2)$ is extracted from
the theoretical value of the leptonic $W^\pm$-decay width $\Gamma_\ell^W$
(rather than from $\mu$-decay) through \cite{DSW}
\bq
  g^2(M_W^2) = 48\pi{{\Gamma_\ell^W}\over {M_W}}
             = {{4\sqrt 2 G_\mu M_W^2}\over{1+\Delta y^{SC}}},
\label{2.4}
\eq
with $\Delta y^{SC} = 3.3\times 10^{-3}$, a constant, practically 
independent of the values of the top-quark mass and Higgs mass,
which in principle enter the radiative one-loop correction 
$\Delta y^{SC}$.\par
     At the one-loop level, the helicity amplitudes depend on twelve 
invariant amplitudes.  In the notation of ref. \cite{FK3S}, 
${\cal H}(\sigma,\lambda,\bar\lambda)$ becomes
\bq
     {\cal H}(\sigma, \lambda,\bar\lambda) =
      S_I^{(\sigma)}{\cal M}_I(\sigma,\lambda,\bar\lambda)+
      S_Q^{(\sigma)}{\cal M}_Q(\sigma,\lambda,\bar\lambda)+
    \sum_{i=2,3,4,6} Y_i^{(\sigma)}(s,t) \bar M_i(\sigma,\lambda,\bar\lambda),
\label{2.5}
\eq
with
\bqa
    S_I^{(\sigma)}(s,t)&=& -tY^{(\sigma)}_5(s,t),\label{2.6} \\
    S_Q^{(\sigma)}(s,t)&=& -{{st}\over{M_Z^2}}Y^{(\sigma)}_5(s,t) 
         +{{s(s-M_Z^2)}\over{M_Z^2}} Y^{(\sigma)}_1(s,t). 
\label{2.7}
\eqa
We refer to ref.~\cite{FK3S} for the explicit definition of the basic 
matrix elements
$\bar M_i(\sigma, \lambda,\bar\lambda)$.  They were chosen 
in such a way that the invariant amplitudes $Y_i^{(\sigma)}(s,t)$
are related to various $s$-channel multipole interactions, a 
$t$-channel-exchange  and a contact interaction.\par
     Out of the set of twelve invariant amplitudes, only 
the three
Born-form invariant amplitudes contain (renormalized) ultraviolet 
divergences and depend on an infrared cut-off.  Accordingly, it was 
suggested in ref. \cite{FK3S} to approximate ${\cal H}(\sigma, \lambda,
\bar\lambda)$ by restricting oneself to an expression of the Born form 
by dropping all other terms in (\ref{2.5}):
\bq
     {\cal H}(\sigma, \lambda,\bar\lambda) =
          S_I^{(\sigma)} {\cal M}_I
                        (\sigma,\lambda,\bar\lambda) \delta_{\sigma,-}
         +S_Q^{(\sigma)}{\cal M}_Q
                        (\sigma,\lambda,\bar\lambda). 
\label{2.8}
\eq
We note at this point that the requirement of a Born-form approximation 
by itself, in general, does not uniquely determine the Born-form invariant
amplitudes $S_I^{(-)}(s,t)$ and $S_Q^{(\pm)}(s,t)$.  A choice of 
basic matrix elements different from $\bar M_i(\sigma, \lambda,\bar\lambda)$,
for $i = 2,3,4,6$, will in general yield different invariant amplitudes,
and in particular also different amplitudes $S_I^{(-)}(s,t)$ and 
$S_Q^{(\pm)}(s,t)$.  One condition for a reasonable Born-form approximation 
is mandatory, however: the basic matrix elements (with the corresponding 
invariant amplitudes) are to be chosen in such a manner that the high-energy 
unitarity constraints on the helicity amplitudes are fulfilled, even upon applying 
the Born-form approximation (\ref{2.8}).  An analysis of the high-energy 
behaviour of the helicity amplitudes (compare \cite{FK3S} and
table A1 in appendix A) reveals that the invariant amplitudes in the
decomposition (\ref{2.5}) fulfil the necessary unitarity constraints
\bq
   S_I^{(\sigma)} \sim S_Q^{(\sigma)}\sim O(1), ~~~~~
   sY_2^{(\sigma)}\sim s Y_3^{(\sigma)}\sim sY_6^{(\sigma)}\sim 
O\left({1\over s}\right),
   ~~~
   sY_4^{(\sigma)}\sim O\left({1\over{s^2}}\right). 
\label{2.9}
\eq
     Dropping the terms with  $i=2,3,4,6$ in (\ref{2.5}) thus yields a 
possible approximation.  For transversely polarized $W$ bosons this 
approximation is numerically successful; the ${\cal M}_I$ term with its
$1/t \sim 1/s(1-\cos\theta)$ forward peak dominates
all other contributions.  For mixed $W$ polarizations
($\lambda=0,\bar\lambda=\pm 1$ and $\lambda=\pm 1,\bar\lambda=0$) 
and for purely longitudinal polarization,
this approximation turns out not to be very satisfactory.\par
     Noting that in the high-energy limit ($\sqrt s \sim  2000$ GeV)
the purely longitudinal amplitude becomes dominant with respect to the mixed 
transverse-longitudinal one \cite{BKRS}
, it is suggestive to modify 
the invariant amplitudes 
(\ref{2.6}),(\ref{2.7}) in the Born form (\ref{2.8}) in such
a manner that the purely longitudinal helicity amplitude is 
reproduced exactly (except for the presence of $S^{(+)}_I$), without 
spoiling the good approximation for the transverse case.
For details we refer to appendix A, and only state the
result.  Choosing $S_I^{(-)}(s,t)$ as in (\ref{2.6}),
but modifying $S_Q^{(\pm)}(s,t)$ in (\ref{2.7}) by adding an appropriate 
contribution containing $Y_2^{(\sigma)}(s,t)$ and 
$Y_6^{(\sigma)}(s,t)$, yields the desired result, i.e.
\bqa
    S_I^{(\sigma)}&=& -tY^{(\sigma)}_5, \label{2.10} \\
    S_Q^{(\sigma)}&=& -{{st}\over{M_Z^2}}Y^{(\sigma)}_5 
         +{{s(s-M_Z^2)}\over{M_Z^2}}  [Y^{(\sigma)}_1
         + {2\over{3-\beta^2}}Y^{(\sigma)}_2 
         + {{\cos^2\theta}\over{3-\beta^2}}Y^{(\sigma)}_6], 
\label{2.11}
\eqa
where $s$ is the centre-of-mass energy squared and 
\bq
    t = M_W^2 - {s\over 2}(1-\beta \cos\theta) ~~~~{\rm and}~~~~~
     \beta = \sqrt { 1- {{4M_W^2}\over s} }.
\label{2.12}
\eq
Using these amplitudes at the one-loop level in the Born  form 
(\ref{2.8}) reproduces the purely longitudinal helicity amplitudes 
(apart from an entirely negligible contribution due to $S^{(+)}_I$)
, and 
at the same time yields an excellent high-energy approximation
for the purely transverse helicities. Numerically this will be 
demonstrated in section 5. \par
    We note that expression  (\ref{2.11}) for $S_Q^{(\pm)}(s,t)$
to be used subsequently differs from the one in ref. \cite{FK3S}, which was 
motivated by unitarity considerations for the mixed polarization.
In terms of $Y_i^{\sigma)}(s,t)$ the previous approximation 
(compare (32) in ref.~\cite{FK3S}) reads
\bqa
   S_I^{(\sigma)}&=& -tY^{(\sigma)}_5, \label{2.13} \\
   S_Q^{(\sigma)}&=& -{{st}\over{M_Z^2}}Y_5^{(\sigma)}
                  +{{s(s-M_Z^2)}\over{M_Z^2}}
                  \left[Y_1^{(\sigma)} + {1\over
2}Y_2^{(\sigma)}-{1\over 2}Y_3^{(\sigma)}
                 -\sigma{1\over 2}{s\over{M_W^2}} \cos\theta
Y_4^{(\sigma)}\right].
\label{2.14}
\eqa
This expression is found to coincide with the so-called 
form-factor approximation (FFA) of ref.~\cite{DBD}
(see also ref.~\cite{D1}). \par
For completeness, in appendix A, we give the 
full expression for the helicity amplitudes, 
when  adopting  (\ref{2.10}), (\ref{2.11}) and (\ref{2.13}), (\ref{2.14}) for 
$S_I^{(\sigma)}(s,t)$ and $S_Q^{(\sigma)}(s,t)$. In appendix B, by evaluating 
$S_I^{(-)}(s,t)$ and $S_Q^{(\pm)}(s,t)$ numerically at the one-loop level, we
compare the different Born-form approximations based on (\ref{2.10}), (\ref{2.11}) 
and on (\ref{2.13}), (\ref{2.14}) for various choices of the $W^+W^-$ 
polarization.  In particular in the high-energy limit
($\sqrt s \cong 2000$ GeV), in which purely longitudinal production
(both $W^+$ and $W^-$ longitudinally polarized) dominates
\cite{BKRS} over
the mixed case (one $W$ transverse, the other one longitudinal),
the novel Born-form approximation (\ref{2.10}), (\ref{2.11}) of the
present paper yields better results than the one based on 
(\ref{2.13}), (\ref{2.14}) examined previously \cite{FK3S},
\cite{DBD}, \cite{D1}.\par
In section 3, we will give 
a simple analytic high-energy approximation for the invariant amplitudes
$S_I^{(-)}(s,t)$  and $S_Q^{(\pm)}(s,t)$ in (\ref{2.10}), (\ref{2.11}). 
This  high-energy approximation will be numerically compared 
with the full one-loop results for 
$S_I^{(-)}(s,t)$ and $S_Q^{(\pm)}(s,t)$
from (\ref{2.10}) and (\ref{2.11}) in section 4.

\medskip

\section{\bf Analytic High-Energy Approximation for 
$S_I^{(-)}(s,t)$ and $S_Q^{(\pm)}(s,t)$}\par
\setcounter{equation}{0}
     The reduction of the set of twelve invariant amplitudes 
to a set of three invariant amplitudes 
in the Born-form approximation provides an important conceptual 
simplification.
The approach develops its full power, however, upon constructing a simple
analytic high-energy approximation for these three invariant amplitudes,
$S_I^{(-)}(s,t)$ and $S_Q^{(\pm)}(s,t)$ at the one-loop level.\par
     
Rescaling the invariant amplitudes by their Born value (\ref{2.1}),
and removing 
the infrared singularities by adding a soft-photon bremsstrahlung 
correction $\delta_{Br}$ \cite{BDSBBK}, we define the amplitudes
$\hat S_I^{(-)}(s,t)$ and $\hat S_Q^{(\pm)}(s,t)$:
\bqa
  \hat S_I^{(-)} &=& {2\over {g^2}}S_I^{(-)} + {1\over 2} \delta_{Br}, 
    \label{3.1}\\
  \hat S_Q^{(\pm)}&=& {1\over {e^2}}S_Q^{(\pm)} + {1\over 2} \delta_{Br}. 
    \label{3.2}
\eqa
In the subsequent discussion of the numerical evaluation of the
invariant amplitudes, we use these hatted quantities, which are 
infrared-finite, but depend on the soft-photon cut-off $\Delta E$.
Note that at tree level, $\hat S_I^{(-)}(s,t) = \hat S_Q^{(\pm)}(s,t)
 = 1$.\par
     In a first step, we approximate $S_I^{(-)}(s,t)$ and $S_Q^{(\pm)}(s,t)$
by taking into account the leading fermion-loop corrections due to the light 
leptons and quarks and the heavy top quark, as well as the initial state
radiation(ISR) in leading-log approximation.\footnote{When supplemented
with the Coulomb correction, this approximation coincides with the 
improved Born approximation (IBA) of refs.~\cite{DBD},\cite{D1}, which
is sufficiently accurate at LEP energies, provided the
correction $\Delta y^{SC}$ from (\ref{2.4}) is introduced, 
\cite{KKS}. Note that
(\ref{3.3}), apart from $\Delta_{LL}$, amounts to nothing else but
using (\ref{2.4}) with $\Delta y^{SC}=0$ for the $SU(2)$ coupling $g$.}
The light fermions
imply replacement of $\alpha(0)$ by the running  electromagnetic 
coupling $\alpha(s)$,
while the top quark yields the well-known SU(2) breaking proportional to 
$m_t^2$.  
The invariant amplitudes become
\bqa  
  \hat S_I^{(-)}  &=& 1+\Delta \alpha(M_W^2)-{{c_W^2}\over{s_W^2}}
                   \Delta\rho + 0.5\Delta_{LL}(s,t),\label{3.3}\\  
  \hat S_Q^{(\pm)}&=& 1+\Delta \alpha(s) +0.5\Delta_{LL}(s,t),
\label{3.4} 
\eqa
with
\bqa
     \Delta \alpha(s)&=& {\alpha\over{3\pi}}\sum_f Q_f^2 
                         \log{s \over {m_f^2}},\label{3.53}\\
        \Delta \rho  &=& {{3g^2}\over{16\pi}}~{{m_t^2}\over{M_W^2}},
                                              \label{3.6}\\
     \Delta_{LL}(s,t)&=&- {{\alpha}\over{\pi}} \Big[ 
                          {3\over 2}\log{{m_e^2}\over s} 
                   + 2\log{{2\Delta E}\over {\sqrt s}}
                \Big(2 + \log {{m_e^2}\over s}  
            +2\log{{M_W^2-u}\over{M_W^2-t}} \nonumber \\ 
            & & ~~~~~~~~~~~~~~~~~~~~~~~~~~~~~~~~~~~~
               +{{s-2M_W^2}\over {s\beta}} \log{{1-\beta}\over{1+\beta}}
                 ~ \Big )\Big ],  
\label{3.7}
\eqa 
where  $\beta$ is given by (\ref{2.12})  and 
\bq
     c_W^2 = 1-s_W^2 = {{M_W^2}\over{M_Z^2}}.
\label{3.8}
\eq
One expects that this approximation, based on fermion loops and the 
leading-log ISR only, will be
insufficient at high energies.  This will be quantitatively discussed in 
section 4.\par
     In order to improve the above approximation (\ref{3.3}), (\ref{3.4}), bosonic
contributions have to be added. As we are aiming at a high-energy 
approximation, we add the high-energy expansions, $S_I^{(-)dom}(s,t)$ 
and $S_Q^{(\pm)dom}(s,t)$, of the one-loop corrections not yet taken into 
account in (\ref{3.3}), (\ref{3.4}) to obtain
\bqa
   \hat S_I^{(-)}  &=& 1+\Delta \alpha(M_W^2)   
                    -{{c_W^2}\over{s_W^2}}\Delta\rho 
                    + 0.5\Delta_{LL}(s,t)+ S_I^{(-)dom}(s,t),\label{3.9}\\
   \hat S_Q^{(\pm)} &=& 1+\Delta \alpha(s) +0.5\Delta_{LL}(s,t)
                    +  S_Q^{(\pm)dom}(s,t).
\label{3.10}
\eqa
For the determination of the explicit form of $S_I^{(-)dom}(s,t)$ 
and $S_Q^{(\pm)dom}(s,t)$ a high-energy approximation of the full
one-loop expressions must be carried out. Fortunately, this task, which is in 
principle straightforward and is in practice time-consuming, can be avoided by
establishing a connection between  $\hat S_I^{(-)}(s,t), \hat S_Q^{(\pm)}(s,t)$
and results given in the literature. In ref. \cite{BDDMS}, without
providing a Born-form representation for the helicity amplitudes, 
the differential cross sections for various polarization states of the
 $W$ bosons were given at one loop in high-energy approximation. \par
The argument which allows one
to connect (linear combinations of) $\hat S_I^{(-)}(s,t)$ and 
$\hat S_Q^{(\pm)}(s,t)$ 
with the results on one-loop cross sections in ref. \cite{BDDMS} is slightly
different for longitudinal and transverse $W$ polarization. As emphasized
before, our choice of basic matrix elements (compare (A.1) to (A.11) and
(\ref{2.10}), (\ref{2.11})) is such that the helicity amplitudes for both
$W^+$ and $W^-$ longitudinally polarized solely depend on
$S_I^{(\pm)}(s,t)$ and $S_Q^{(\pm)}(s,t)$. Accordingly, by comparing the
differential cross sections in terms of $\hat S_I^{(\pm)}(s,t)$ and 
$\hat S_Q^{(\pm)}(s,t)$ with the high-energy approximation for the differential 
cross sections at the one-loop level from ref. \cite{BDDMS}, we obtain
\bqa  
  2s_W^2\hat S_Q^{(-)} +(2c_W^2-1)\hat S_I^{(-)} &=& 1 + {1\over 2}
               [C_{-,L}^B +C_{-,L}^F], \label{3.11} \\
  \hat S_Q^{(+)} + {{2c_W^2-1}\over{2s_W^2}}\hat S_I^{(+)} &=& 1+{1\over 2}
               [C_{+,L}^B +C_{+,L}^F],  
\label{3.12}
\eqa
where the high-energy one-loop corrections to the cross sections $C_{-,L}^B (s,t)$
etc., are given in ref. \cite{BDDMS} and will be reproduced below. For details
on the derivation of (\ref{3.11}), (\ref{3.12}) and (\ref{3.13}) below, we
refer to appendix C.
\par
For transversely polarized $W$ bosons, the helicity amplitudes in general depend
on additional  invariant amplitudes besides $S^{(\pm)}_I(s,t)$ 
and $S^{(\pm)}_Q(s,t)$. Nevertheless,
from the numerical analysis of appendix B, we know that the Born-form 
approximation is excellent at high energies when inserting one-loop results
for the invariant amplitudes. In fact, only $S_I^{(-)}(s,t)$  appears in
the high-energy approximation, thus implying (compare appendix C)
\bq
   \hat S_I^{(-)}= 1+ {1\over 2} 
               [C_{-,T}^B +C_{-,T}^F].  \label{3.13}
\eq
  In order to obtain explicit expressions for $\hat S_I^{(-)}$ and
$\hat S_Q^{(\pm)}$ in the high-energy approximation, we now 
neglect the extremely small\footnote{The typical magnitude of
$\hat S_I^{(+)}$ is of order $10^{-4}$.} contribution from 
$\hat S_I^{(+)}$ in (\ref{3.12}).  The dominant contributions to the
coefficients $C_{\pm, L}^{B/F}$, $C_{-,T}^{B/F}$ depend 
(quadratically or linearly) on log($s$) and are given in ref. \cite{BDDMS}
\footnote{Compare (12) of ref. \cite{BDDMS}.}.  Substitution on the
right-hand side in (\ref{3.11}) to (\ref{3.13}) and numerical
evaluation shows that the main part of the full one-loop results for
$\hat S_I^{(-)}$ and $\hat S_Q^{(\pm)}$  is reproduced by these
leading terms in the cross-section coefficients $C_{\pm, L}^{B/F}$ and 
$C_{-,T}^{B/F}$.  In fact, the difference between this high-energy
approximation and the full one-loop results for 
$\hat S_I^{(-)}$ and $\hat S_Q^{(\pm)}$  
is only weakly dependent on energy and on the mass of the Higgs boson for 
reasonable Higgs masses of $M_H=$ 100 GeV to $M_H=$ 300 GeV. 
Accordingly, the rest terms beyond 
the leading order may be taken care of by adding small constants, correctly 
adjusted in magnitude to approximate the full one-loop results for 
$\hat S_I^{(-)}$ and $\hat S_Q^{(\pm)}$ wherein, for definiteness, 
the energy is chosen as $\sqrt{s}=2$ TeV, and the Higgs mass as 
$M_H$ = 200 GeV.  In this way, we arrive at 
the following results: 
\bqa
   S_I^{(-)dom}
     &= & {{\alpha}\over{4s_W^2}}
     \Biggl[-{{1+2c_W^2+8c_W^4}\over{4c_W^2}}(\log{s\over{M_W^2}})^2
       + (4+2{s\over u})(\log{s\over{M_W^2}})(\log{s\over t}) \nonumber \\
     & &-({{s[s(1-6c_W^2)+3t]}\over{4c_W^2(t^2+u^2)}}+
          {{s(1-6c_W^2)}\over {2c_W^2 u}}) (\log{s\over t})^2 \nonumber \\
     & & -{{3st}\over{2(t^2+u^2)}}(\log{s\over u})^2 
         -{{2s}\over u}(\log{s\over t})(\log{s\over u}) \nonumber \\
     & &+{{3(s_W^4+3c_W^4)}\over{4c_W^2}}\log{s\over {M_W^2}}
        - {{1-4c_W^2+8c_W^4}\over{2c_W^2}}(\log{s\over{M_W^2}})(\log c_W^2)
           \nonumber \\
     & & + 2(1-2c_W^2)(\log{t\over u})(\log {s\over{M_Z^2}}) 
          -2s_W^2(\log {t \over u})^2 -8Sp(-{u\over t}) \nonumber \\
     & & -{{s[3s+t+6c_W^2(s+3t)]}\over{4c_W^2(t^2+u^2)}}\log{s\over t}
         -{{(1-6c_W^2)su}\over{4c_W^2(t^2+u^2)}} \Biggr] - 0.012,  \label{3.14}
\eqa
\bqa
     S_Q^{(-)dom} &=& {{\alpha}\over{8\pi s_W^2}}
    \Biggl [-{{3-4c_W^2+12c_W^4-16c_W^6}\over{4c_W^2s_W^2}}(\log{s\over{M_W^2}})^2 
          \nonumber \\
    & &+ {{56-57c_W^2+36c_W^4-36c_W^6}\over{6c_W^2s_W^2}}\log{s\over{M_W^2}}
         \nonumber \\
     & &- (1-2c_W^2){{2(1-2c_W^2)^2+1}\over{2c_W^2s_W^2}}\log c_W^2 
          \log {s\over{M_W^2}}
        + (4 + 2{{1-2c_W^2}\over{s_W^2}}~{s\over u})\log{s\over{M_W^2}}
           \log{s\over t} \nonumber \\
     & &+{{(1-2c_W^2)^3}\over{c_W^2s_W^2}}(\log {u\over t})(\log{s\over{M_Z^2}})
        -2{{1-2c_W^2}\over{s_W^2}}~{s\over u}(\log{s\over t})(\log{s\over u})
            \nonumber \\
     & &-\Big[ {{1-16c_W^2+20c_W^4}\over{4c_W^2s_W^2}}~{s\over u}
            +{{1-2c_W^2}\over{4c_W^2s_W^2}}s{{s+3t-6c_W^2s}
\over{t^2+u^2}}\Big]
              (\log{s\over t})^2 \nonumber \\
     & & -( {1\over{4c_W^2s_W^2}}~{s\over t} 
            +{{1-2c_W^2}\over{2s_W^2}}~{{3st}\over{t^2+u^2}} )(\log{s\over u})^2
            \nonumber \\
     & & -4s_W^2(\log{u\over t})^2 -16s_W^2 Sp(-{u\over t}) 
         -{{1-2c_W^2}\over{4c_W^2s_W^2}} s{{3s+t+6c_W^2(s+3t)}\over{t^2+u^2}}
           \log{s\over t} \nonumber \\
     & & -{{(1-2c_W^2)(1-6c_W^2)}\over{4c_W^2s_W^2}}~{{su}\over{t^2+u^2}} 
         + {3\over 2}~ {{m_t^2}\over{s_W^2M_W^2}} 
\log{{m_t^2}\over s} \Biggr]
         +0.030, \label{3.15} 
\eqa
\bqa
     S_Q^{(+)dom} &=& {{\alpha}\over{4\pi}}
      \Biggl[ -{{5s_W^4+3c_W^4}\over{4c_W^2s_W^2}} (\log{s\over {M_W^2}})^2
          + {{65s_W^2+18c_W^4}\over{6c_W^2s_W^2}}\log{s\over {M_W^2}}
          \nonumber \\
     & & - {{(1-2s_W^2)^2}\over{2c_W^2s_W^2}}\log c_W^2~\log{s\over{M_W^2}}
         + 2{{1-2c_W^2}\over{c_W^2}}\log{u\over t} \log{s\over{M_Z^2}}
         +{s\over{2c_W^2 u}}(\log{s\over t})^2  \nonumber \\
     & & -{s\over{2c_W^2 t}}(\log{s\over u})^2 -2(\log{u\over t})^2 
         -8Sp(-{u\over t}) + {{3m_t^2}\over{2s_W^2 M_W^2}}\log{{m_t^2}\over s}
         \Biggr] + 0.045. \label{3.16}
\eqa
\par
     As mentioned, the constants in (\ref{3.14}) to (\ref{3.16}) are 
adjusted such that the full one-loop results for $\hat S_I^{(-)}$ and
$\hat S_Q^{(\pm)}$ evaluated  for $M_H=200$ GeV at $\sqrt s = 2$ TeV 
are well reproduced.  Table 1 shows the values of the constants,
if these are  adjusted  to $\hat S_I^{(-)}$ and $\hat S_Q^{(\pm)}$
evaluated at different values of $M_H$ and $\sqrt s$.
Table 1 demonstrates that the asymptotic region for $S_I^{(-)}$ and 
$S_Q^{(\pm)}$ is indeed reached
at $\sqrt s>2$ TeV, the dependence on $M_H$ being weak.
Using constants determined at lower energy may be useful, if a better 
approximation at lower energy is desired, thus effectively
taking into account non-leading contributions of order $1/s$
(compare $S_Q^{(+)}$, in particular).

\begin{table}[htbp]\centering
\footnotesize{
\begin{tabular}{|r|r|r|r|}\hline
     $\sqrt s$ &   500 GeV       &   2 TeV      &   20 TeV  \\ \hline
     $M_H$(GeV)& \multicolumn{3}{c|}{$S_I^{(-)dom}(s,t)$} \\ \hline
     100       & $-$0.012  &  $-$0.014   & $-$0.014 \\ \hline
     200       & $-$0.009  &  $-$0.012   & $-$0.011 \\ \hline
     300       & $-$0.007  &  $-$0.009   & $-$0.009 \\ \hline\hline
     $M_H$(GeV)& \multicolumn{3}{c|}{$S_Q^{(-)dom}(s,t)$} \\ \hline
     100       &   0.024   &  0.024      &  0.024  \\ \hline
     200       &   0.030   &  0.030      &  0.030  \\ \hline
     300       &   0.035   &  0.035      &  0.035  \\ \hline\hline
     $M_H$(GeV)& \multicolumn{3}{c|}{$S_Q^{(+)dom}(s,t)$} \\ \hline
     100       &  0.059    &  0.041      &  0.039  \\ \hline
     200       &  0.063    &  0.045      &  0.042  \\ \hline
     300       &  0.065    &  0.046      &  0.044  \\ \hline
\end{tabular}

\medskip
\caption{ The constants in $S_I^{(-)dom}(s,t)$ and
$S_Q^{(\pm)dom}(s,t)$ adjusted  such that (\ref{3.14}) to (\ref{3.16})
yield a good approximation of the full one-loop
result for  $\hat S_I^{(-)}(s,t)$ and $\hat S_Q^{(\pm)}(s,t)$
at different Higgs masses and energies.} }
\label{Table 1}
\end{table}

\vskip 0.8 truecm 

\section{\bf  Numerical Evaluation of $\hat S_I^{(-)}(s,t)$ and 
$\hat S_Q^{(\pm)}(s,t)$}\par
\setcounter{equation}{0}
     In order to compare the various approximations for
$\hat S_I^{(-)}(s,t)$
and $\hat S^{(\pm)}_Q(s,t)$ of section 3, we numerically evaluated 
these for the choice of two energies, $\sqrt s = 2000$ GeV and $\sqrt s =500$ GeV.  
\par
As basic electroweak input parameters, we use the fine-structure constant
$\alpha$ and the $Z^0$ and $W^\pm$ masses $M_Z$ and $M_W$. Accordingly,
the coupling constants $e$ and $g$ are given by
\bq
     e = \sqrt{4\pi\alpha} = 0.3028, ~~~~~ 
     g = {e\over {s_W}} = e{{M_Z}\over {\sqrt { M_Z^2-M_W^2}} }~,
\eq
with \cite{Kim}
\bq
     M_Z=91.186~{\rm GeV}, ~~~~~~M_W = 80.430~{\rm GeV}.
\eq
As discussed in section 3, the results are insensitive to the exact value of $M_H$, 
which is chosen as $M_H=200$ GeV. Finally, we specify lepton and quark masses

\bqa
     m_u = 0.041~{\rm GeV}, &  m_d =0.041~{\rm GeV},   &  m_c = 1.5~{\rm GeV}, \nonumber\\  
     m_s = 0.15~{\rm GeV},  &  m_t = 175.6~{\rm GeV},  &  m_b = 4.5~{\rm GeV}, 
\eqa
and the soft-photon cut-off
\bq
     \Delta E=0.025 \sqrt s.
\eq
The numerical evaluation was carried out in three steps:\par
i) the full one-loop results (including soft-photon radiation)
for $S_I^{(-)}(s,t)$ and $S_Q^{(\pm)}(s,t)$ 
from (\ref{2.10}), (\ref{2.11}) were numerically produced
by evaluating the computer code created by one of the authors
\cite{K1};\footnote{The results of ref.~\cite{K1} were verified to agree
with refs.~\cite{BDSBBK}, \cite{DBD}, \cite{D1}.} \par
  ii) the fermion-loop approximation with ISR, (\ref{3.3}), (\ref{3.4}), 
was numerically evaluated;
\par
   iii) the full high-energy approximation
based on (\ref{3.9}),(\ref{3.10}) 
and (\ref{3.14}) to (\ref{3.16}) was
numerically evaluated.\par
   The results of the numerical analysis are presented in 
fig. 1 for $S_I^{(-)}(s,t)$ and in figs. 2 and 3 for 
$S_Q^{(\pm)}(s,t)$.\par
     Figure 1 shows remarkably good agreement of the high-energy
approximation with the full one-loop result for $\hat S_I^{(-)}$,  
apart from a small deviation in the backward region at $\sqrt s= 500$ GeV.
According to figs. 2 and 3, also $\hat S_Q^{(\pm)}(s,t)$ at $\sqrt s= 2000$ GeV
is very well represented by the high-energy approximation, while at
$\sqrt s= 500$ GeV, there is some departure from the full one-loop result.
The discrepancy between  the high-energy approximation and  the 
full one-loop result for  $\hat S_Q^{(+)}$ at $\sqrt s=500$ GeV (see
fig. 3)
can be removed by using  a constant of magnitude  +0.063, slightly larger than
the value of 0.045 that gives a good approximation at $\sqrt s= 2000$ GeV.
Compare table 1. \par
\begin{figure}[htbp]\centering
\epsfysize=18cm
\centerline{\epsffile{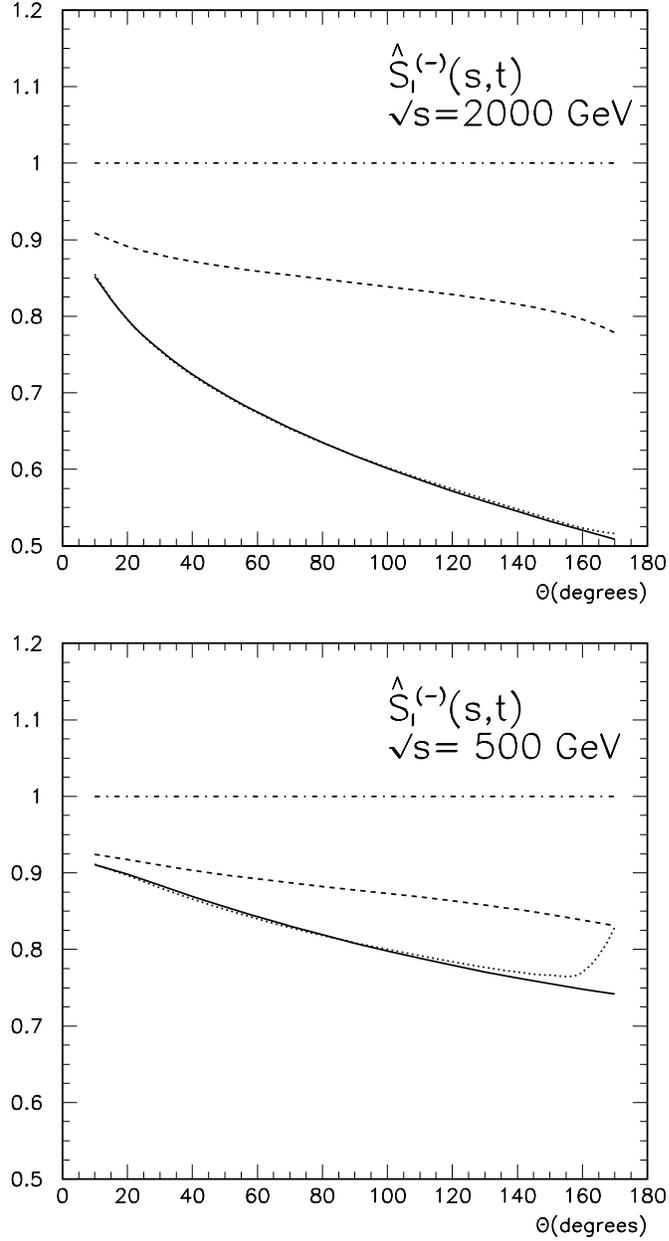}}
\caption{ The Born-form invariant amplitude $\hat S_I^{(-)}(s,t)$
as a function of the W production angle, $\theta$, for
$\sqrt{s}=2000$ GeV and $\sqrt{s}=500$ GeV in
(i) the full one-loop evaluation including soft-photon bremsstrahlung 
(solid line), 
(ii) the fermion-loop approximation including soft-photon 
bremsstrahlung (dashed line),
(iii) the high-energy approximation based on (3.9),(3.10) and 
(3.14) to (3.16)(dotted line), 
(iv) the Born approximation (dash-dotted line).}
\label{Fig. 1}
\end{figure}

\begin{figure}[htbp]\centering
\epsfysize=18cm
\centerline{\epsffile{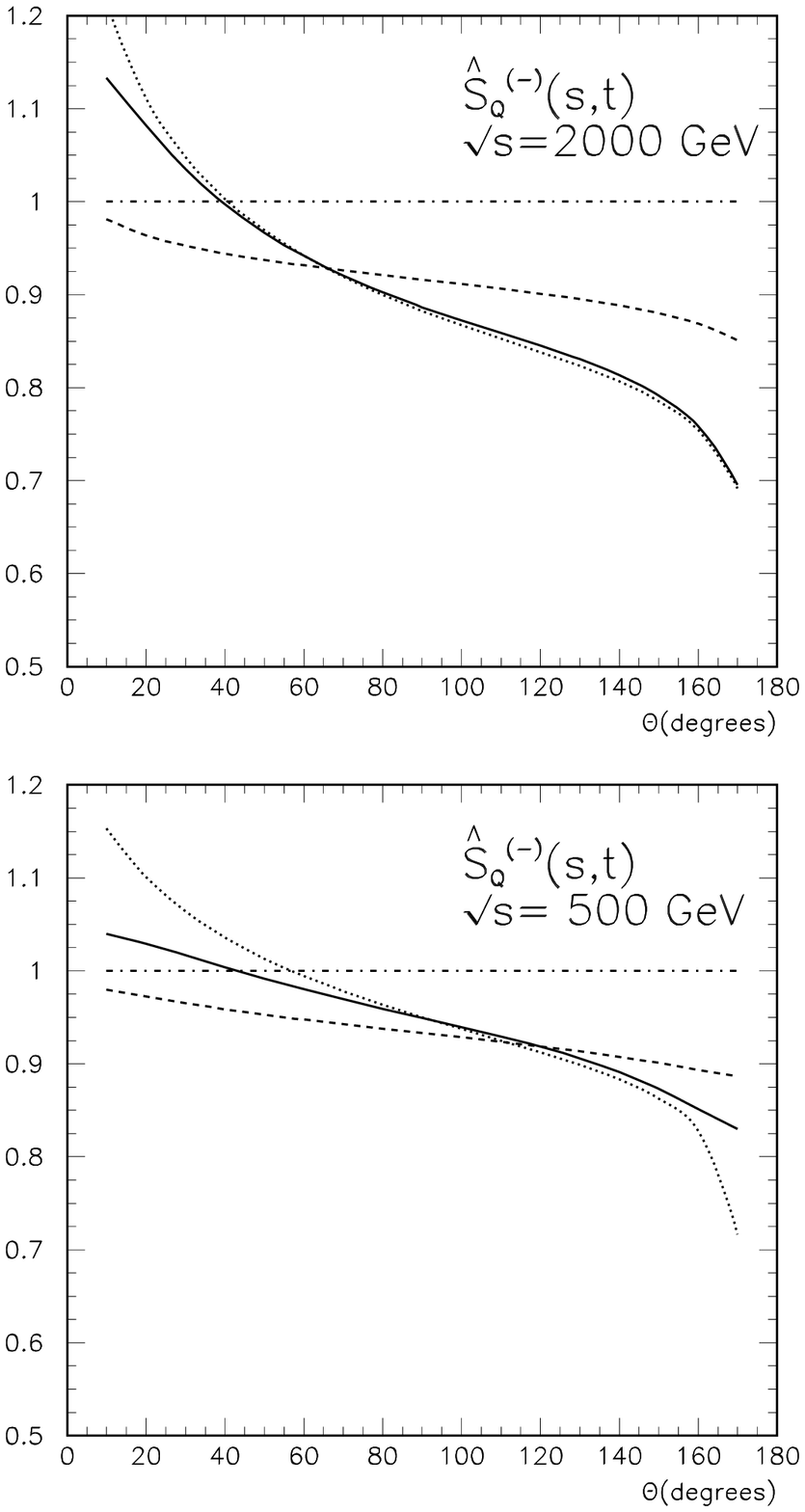}}
\caption{ Same as fig.1, but for $\hat S_Q^{(-)}(s,t)$}
\label{Fig. 2}
\end{figure}

\begin{figure}[htbp]\centering
\epsfysize=18cm
\centerline{\epsffile{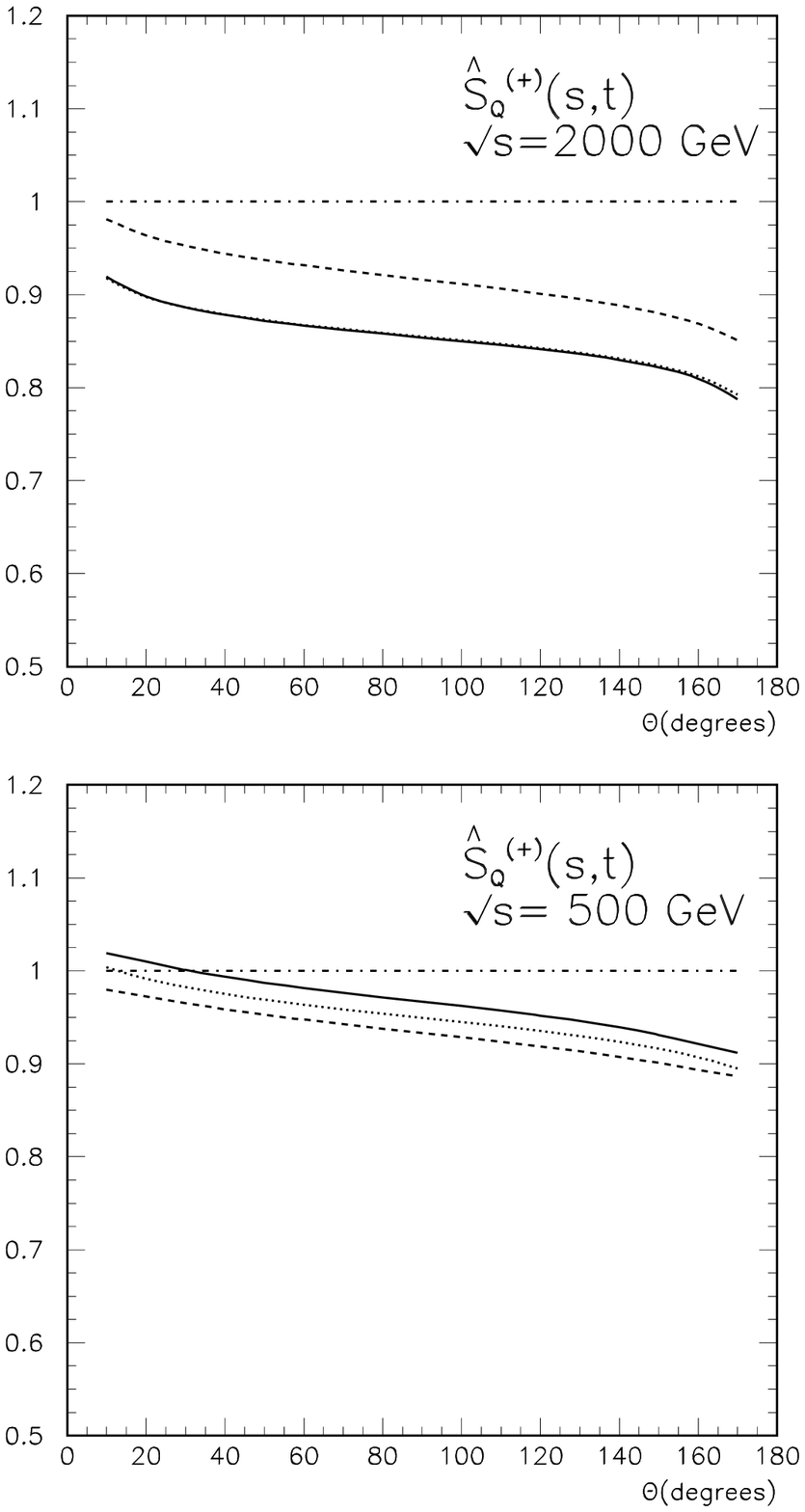}}
\caption{ Same as fig.1, but for $\hat S_Q^{(+)}(s,t)$}
\label{Fig. 3}
\end{figure}
\newpage

     It is  worth stressing the large difference in
$\hat S_I^{(-)}(s,t)$ 
and $\hat S_Q^{(\pm)}(s,t)$, particularly at 2000 GeV, between the 
fermion-loop approximation with leading-log ISR, (\ref{3.3}),(\ref{3.4}) 
and the full one-loop
result or, equivalently, the high-energy approximation based on (\ref{3.9})
and (\ref{3.10}) upon substituting (\ref{3.14}) to (\ref{3.16}).
As the soft-photon cut-off $\Delta E$ drops out of the difference
between the fermion-loop approximation with leading-log ISR 
and the full one-loop results
(or the high-energy approximation), the large difference between 
the corresponding  curves  in figs.~1 to 3 constitutes a genuine effect of virtual 
electroweak radiative corrections.  This effect of, e.g., 10$\%$ to
30$\%$ in $\hat S_I^{(-)}(s,t)$ at $\sqrt s=2000$ GeV, is independent
of the Higgs-boson mass, but it constitutes a genuine effect
due to the non-Abelian form of the electroweak theory, which gives
rise to various box diagrams and vertex corrections of bosonic origin.
\par
     In section 2 and appendix A, we pointed out that 
$S_I^{(-)}(s,t)$ and $S_Q^{(\pm)}(s,t)$ from  (\ref{2.10}), (\ref{2.11}),
evaluated at one-loop level, yield an excellent (Born-form)
approximation of the differential cross sections for various $W$
polarizations in the high-energy limit. Since 
$S_I^{(-)}(s,t)$ and $S_Q^{(\pm)}(s,t)$ at one loop
are well approximated 
by the simple high-energy form (\ref{3.9}), (\ref{3.10}),
we expect excellent results for differential cross sections when
employing (\ref{3.9}), (\ref{3.10}).
The detailed numerical investigation of various differential cross sections 
is the subject of section 5.

\vskip 0.8 truecm

\section{Numerical Results for Differential Cross Sections,
including $W$-Polarization}
\setcounter{equation}{0}

     In this section, we will present a detailed comparison of
the numerical results obtained for various differential cross sections 
at $\sqrt s=2000$ GeV and at $\sqrt s=500$ GeV in \par
   i) the full one-loop evaluation, generated by the computer code of
ref. \cite{K1}, \par
  ii) the Born-form approximation, using the exact one-loop 
expressions for $S_I^{(-)}(s,t)$ and $S_Q^{(\pm)}(s,t)$ generated by the
computer code of ref. \cite{K1}, \par
 iii) the high-energy Born-form approximation with $S_I^{(-)}(s,t)$ and 
$S_Q^{(\pm)}(s,t)$ evaluated in the high-energy approximation 
(\ref{3.9}), (\ref{3.10}) upon substituting (\ref{3.14}) to (\ref{3.16}). \par
     The accuracy of a specific approximation will be quantified by the
percentage deviation
\bq
   \Delta(\%) \equiv {{d\sigma_{approx.} -d\sigma_{full~one-loop}}\over
                      {d\sigma_{Born}}}, \label{5.1}
\eq
which is obviously independent of the arbitrary soft-photon 
cut-off $\Delta E$. \par  

\begin{table}[htbp]\centering
{\footnotesize
\begin{tabular}{|c||l|r||l|r||l|l|}\hline
         & \multicolumn{2}{c||}{High-energy Born-form} 
         & \multicolumn{2}{c||}{Born-form}
         & \multicolumn{1}{c|}{Full~one-loop}
         & \multicolumn{1}{c|}{Born} \\
         & \multicolumn{2}{c||}{approximation}
         & \multicolumn{2}{c||}{approximation}
         &    &   \\ \cline {2-7}
         & \multicolumn{1}{c|}{$\sigma$(pb)}
         & \multicolumn{1}{c||}{$\Delta(\%)$}
         & \multicolumn{1}{c|}{$\sigma$(pb)}
         & \multicolumn{1}{c||}{$\Delta(\%)$}
         & \multicolumn{1}{c|}{$\sigma$(pb)}
         & \multicolumn{1}{c|}{$\sigma$(pb)}
            \\  \hline \hline
     & \multicolumn {6}{c|}{$\sqrt s = 2000$ GeV } \\ \hline
  ``Unpol.''
     & 1.461$\times 10^{-1}$ & +0.14 & 1.461$\times 10^{-1}$  &  +0.16 
     & 1.457$\times 10^{-1}$         & 2.758$\times 10^{-1}$ \\ \hline 
  Transv.
     & 1.422$\times 10^{-1}$ & +0.19 & 1.423$\times 10^{-1}$  &  +0.19 
     & 1.417$\times 10^{-1}$         & 2.683$\times 10^{-1}$ \\ \hline 
  Longit.
     & 3.524$\times 10^{-3}$ &$-$0.13& 3.533$\times 10^{-3}$  &  0.00 
     & 3.533$\times 10^{-3}$         & 6.788$\times 10^{-3}$ \\ \hline 
  Mixed
     & 2.911$\times 10^{-4}$&$-$14.80& 2.909$\times 10^{-4}$ &$-$14.83 
     & 3.833$\times 10^{-4}$         & 6.229$\times 10^{-4}$ \\ \hline 
       \hline
     & \multicolumn {6}{c|}{$\sqrt s = 500$ GeV } \\ \hline
  ``Unpol.''
     & 3.448 &$-$0.42& 3.462 &$-$0.11 & 3.467  & 4.545 \\ \hline 
  Transv.
     & 3.260 &$-$0.34& 3.274 &$-$0.01 & 3.274  & 4.294 \\ \hline 
  Longit.
     & 8.284$\times 10^{-2}$ &$-$0.36& 8.323$\times 10^{-2}$  &  0.00 
     & 8.323$\times 10^{-2}$         & 1.091$\times 10^{-1}$ \\ \hline 
  Mixed
     & 1.033$\times 10^{-1}$ &$-$3.19& 1.034$\times 10^{-1}$  &$-$3.13 
     & 1.078$\times 10^{-1}$         & 1.419$\times 10^{-1}$ \\ \hline 
\end{tabular} 

\medskip
\caption{ The total  cross section  for  $W$ pair production 
(obtained by integration over the angular range of the production angle of
$10^\circ \leq \vartheta \leq 170^\circ$)
at $\sqrt s= 2000$ GeV and $\sqrt s= 500$ GeV. Rows show the 
results when summing over the $W^+W^-$ spins (``unpol.'') and the results for
the various cases of polarization of the produced $W^+$ and $W^-$. 
The first column shows the result of
the Born-form approximation using the high-energy approximation
for $S_I^{(-)}$ and $S_Q^{(\pm)}$ given by (\ref{3.9})
and (\ref{3.10}) with (\ref{3.14}) to (\ref{3.16}).  
The second column gives the result of the
Born-form approximation obtained  by
evaluating (\ref{2.10}) and (\ref{2.11}) at the one-loop level exactly. The
third column shows the full one-loop result and the Born approximation. 
}  }
\label{Table 2}
\end{table}
\vspace {0.5cm}
     First of all, we concentrate on the total cross sections for fixed
polarization of the produced $W$ bosons.  Table 2  shows, for $\sqrt s=2000$ GeV 
and $\sqrt s=500$ GeV, the cross section summed over the polarization of the 
$W$ bosons (``unpolarized") as well as the cross sections for the cases where 
both the $W^+$ and the $W^-$ are transversely and longitudinally polarized, and, 
finally, the cross section for the case of one longitudinally and 
one transversely polarized $W$ 
(``mixed" ).
One observes that there is no significant difference between the 
high-energy Born-form approximation (iii) and the Born-form approximation (ii)
in all these cases. For the unpolarized, transverse and longitudinal cases,
the accuracy of the Born-form appproximation, with $\Delta < 0.5 \%$, is 
truly excellent. As the mixed polarization at $\sqrt s=2000$ GeV 
contributes only a tiny fraction of about 2 per mille to the cross section, 
the fairly large deviation between
the Born-form approximation and the full one-loop results is
irrelevant with respect to future experiments.  At $\sqrt s=500$ GeV, 
the mixed case contributes about 3\% percent to the cross section, 
and an accuracy of 3\% is therefore sufficient.\par

\begin{table}[htbp]\centering
{\footnotesize
\begin{tabular}{|r||l|r||l|r||l|l|}\hline
   angle & \multicolumn{2}{c||}{High-energy Born-form}
         & \multicolumn{2}{c||}{Born-form}
         & \multicolumn{1}{c|}{Full one-loop}
         & \multicolumn{1}{c|}{Born} \\
 {$(^\circ)$}   & \multicolumn{2}{c||}{approximation}
         & \multicolumn{2}{c||}{approximation}
         &  &  \\ \hline
         & \multicolumn{1}{c|}{${{d\sigma}\over{d\cos\theta}}$(pb)}
         & \multicolumn{1}{c||}{$\Delta(\%)$}
         & \multicolumn{1}{c|}{${{d\sigma}\over{d\cos\theta}}$(pb)}
         & \multicolumn{1}{c||}{$\Delta(\%)$}
         & \multicolumn{1}{c|}{${{d\sigma}\over{d\cos\theta}}$(pb)}
         & \multicolumn{1}{c|}{${{d\sigma}\over{d\cos\theta}}$(pb)}
            \\  \hline \hline
     & \multicolumn {6}{c|}{$\sigma_{``unpol''}$ } \\ \hline
  10 & 3.785 & +0.66 & 3.751    &  +0.02 & 3.750     & 5.329 \\ \hline 
  30 & 2.588$\times 10^{-1}$ & +0.04 & 2.599$\times 10^{-1}$  &  +0.26 
     & 2.586$\times 10^{-1}$         & 5.070$\times 10^{-1}$ \\ \hline 
  50 & 5.511$\times 10^{-2}$ & +0.08 & 5.551$\times 10^{-2}$  &  +0.37 
     & 5.500$\times 10^{-2}$         & 1.382$\times 10^{-1}$ \\ \hline 
  70 & 1.717$\times 10^{-2}$ & +0.02 & 1.727$\times 10^{-2}$  &  +0.21 
     & 1.716$\times 10^{-2}$         & 5.208$\times 10^{-2}$ \\ \hline 
  90 & 7.570$\times 10^{-3}$ &$-$0.08& 7.565$\times 10^{-3}$  &$-$0.10 
     & 7.592$\times 10^{-3}$         & 2.605$\times 10^{-2}$ \\ \hline 
 110 & 4.104$\times 10^{-3}$ &$-$0.19& 4.074$\times 10^{-3}$  &$-$0.38 
     & 4.135$\times 10^{-3}$         & 1.599$\times 10^{-2}$ \\ \hline 
 130 & 2.034$\times 10^{-3}$ &$-$0.20& 2.007$\times 10^{-3}$  &$-$0.49 
     & 2.053$\times 10^{-3}$         & 9.466$\times 10^{-3}$ \\ \hline 
 150 & 6.720$\times 10^{-4}$ & +0.11 & 6.589$\times 10^{-4}$  &$-$0.22 
     & 6.678$\times 10^{-4}$         & 3.993$\times 10^{-3}$ \\ \hline 
 170 & 8.182$\times 10^{-5}$ & +2.79 & 7.567$\times 10^{-5}$  &  +1.76 
     & 6.520$\times 10^{-5}$         & 5.961$\times 10^{-4}$ \\ \hline 
       \hline
      & \multicolumn {6}{c|}{$\sigma_T $ } \\ \hline
  10 & 3.784 & +0.66 & 3.751     &  +0.04 & 3.749     & 5.329 \\ \hline
  30 & 2.574$\times 10^{-1}$ & +0.08  & 2.585$\times 10^{-1}$  & +0.30 
     & 2.570$\times 10^{-1}$          & 5.050$\times 10^{-1}$ \\ \hline
  50 & 5.290$\times 10^{-2}$ & +0.15  & 5.330$\times 10^{-2}$  & +0.45 
     & 5.270$\times 10^{-2}$          & 1.347$\times 10^{-1}$ \\ \hline
  70 & 1.445$\times 10^{-2}$ & +0.17  & 1.454$\times 10^{-2}$  & +0.36 
     & 1.437$\times 10^{-2}$          & 4.720$\times 10^{-2}$ \\ \hline
  90 & 4.891$\times 10^{-3}$ & +0.07  & 4.875$\times 10^{-3}$  &$-$0.00 
     & 4.876$\times 10^{-3}$          & 2.072$\times 10^{-2}$ \\ \hline
 110 & 1.998$\times 10^{-3}$ &$-$0.11 & 1.956$\times 10^{-3}$  &$-$0.48 
     & 2.010$\times 10^{-3}$          & 1.135$\times 10^{-2}$ \\ \hline
 130 & 7.747$\times 10^{-4}$ &$-$0.25 & 7.400$\times 10^{-4}$  &$-$0.80 
     & 7.907$\times 10^{-4}$          & 6.367$\times 10^{-3}$ \\ \hline
 150 & 1.837$\times 10^{-4}$ &$-$0.19 & 1.689$\times 10^{-4}$  &$-$0.76 
     & 1.887$\times 10^{-4}$          & 2.603$\times 10^{-3}$\\ \hline
 170 & 9.944$\times 10^{-6}$ & +1.18  & 5.448$\times 10^{-6}$  &$-$0.26 
     & 6.262$\times 10^{-6}$          & 3.124$\times 10^{-4}$ \\
          \hline \hline
      & \multicolumn {6}{c|}{$\sigma_L $} \\ \hline
  10 & 4.644$\times 10^{-5}$  & +7.71  & 4.329$\times 10^{-5}$  & 0.00 
     & 4.329$\times 10^{-5}$           & 4.084$\times 10^{-5}$ \\ \hline
  30 & 8.995$\times 10^{-4}$  & +0.88  & 8.894$\times 10^{-4}$  & 0.00 
     & 8.894$\times 10^{-4}$           & 1.146$\times 10^{-3}$ \\ \hline
  50 & 1.917$\times 10^{-3}$  & +0.07  & 1.915$\times 10^{-3}$  & 0.00 
     & 1.915$\times 10^{-3}$           & 2.917$\times 10^{-3}$ \\ \hline
  70 & 2.574$\times 10^{-3}$  & $-$0.13& 2.580$\times 10^{-3}$  & 0.00 
     & 2.580$\times 10^{-3}$           & 4.487$\times 10^{-3}$ \\ \hline
  90 & 2.614$\times 10^{-3}$  & $-$0.21& 2.625$\times 10^{-3}$  & 0.00 
     & 2.625$\times 10^{-3}$           & 5.126$\times 10^{-3}$ \\ \hline
 110 & 2.074$\times 10^{-3}$  & $-$0.24& 2.085$\times 10^{-3}$  & 0.00 
     & 2.085$\times 10^{-3}$           & 4.546$\times 10^{-3}$ \\ \hline
 130 & 1.229$\times 10^{-3}$  & $-$0.23& 1.236$\times 10^{-3}$  & 0.00 
     & 1.236$\times 10^{-3}$           & 3.028$\times 10^{-3}$ \\ \hline
 150 & 4.512$\times 10^{-3}$  & $-$0.12& 4.528$\times 10^{-4}$  &+0.01 
     & 4.527$\times 10^{-4}$           & 1.292$\times 10^{-3}$ \\ \hline
 170 & 3.925$\times 10^{-5}$  &  +0.55 & 3.834$\times 10^{-5}$  &$-$0.04 
     & 3.840$\times 10^{-5}$           & 1.559$\times 10^{-4}$ \\ 
          \hline \hline
    & \multicolumn {6}{c|} {$\sigma_{mixed}$} \\ \hline
 10 & 6.644$\times 10^{-4}$ &$-$36.56 & 6.425$\times 10^{-4}$ &$-$39.22 
    & 9.655$\times 10^{-4}$           & 8.236$\times 10^{-4}$ \\ \hline
 30 & 4.956$\times 10^{-4}$ &$-$19.56 & 4.946$\times 10^{-4}$ &$-$19.68 
    & 6.608$\times 10^{-4}$           & 8.444$\times 10^{-4}$ \\ \hline
 50 & 2.948$\times 10^{-4}$ &$-$15.15 & 2.959$\times 10^{-4}$ &$-$14.97 
    & 3.907$\times 10^{-4}$           & 6.332$\times 10^{-4}$ \\ \hline
 70 & 1.493$\times 10^{-4}$ &$-$13.83 & 1.500$\times 10^{-4}$ &$-$13.65
    & 2.037$\times 10^{-4}$           & 3.933$\times 10^{-4}$ \\ \hline
 90 & 6.524$\times 10^{-5}$ &$-$13.14 & 6.528$\times 10^{-5}$ &$-$13.12
    & 9.151$\times 10^{-5}$           & 1.999$\times 10^{-4}$  \\ \hline
110 & 3.254$\times 10^{-5}$ & $-$8.12 & 3.252$\times 10^{-5}$ &$-$10.57 
    & 4.010$\times 10^{-5}$           & 9.307$\times 10^{-5}$ \\ \hline
130 & 3.065$\times 10^{-5}$ &   +5.86 & 3.082$\times 10^{-5}$ &  +6.10 
    & 2.649$\times 10^{-5}$           & 7.099$\times 10^{-5}$ \\ \hline
150 & 3.707$\times 10^{-5}$ &  +10.89 & 3.723$\times 10^{-5}$ & +11.05 
    & 2.638$\times 10^{-5}$           & 9.818$\times 10^{-5}$ \\ \hline
170 & 3.262$\times 10^{-5}$ &   +9.45 & 3.188$\times 10^{-5}$ &  +8.87 
    & 2.054$\times 10^{-5}$           & 1.278$\times 10^{-4}$ \\ \hline
\end{tabular} 
\par

\medskip
\caption{ The differential cross section 
for $e^+e^-\to W^+W^-$ at $\sqrt s = 2000$ GeV for various polarization
states of the $W^+$ and $W^-$. 
The first column shows the 
Born-form approximation, using the high-energy approximation
for $S_I^{(-)}(s,t)$ and $S_Q^{(\pm)}(s,t)$ from (\ref{3.9}), (\ref{3.10}) 
with (\ref{3.14}) to (\ref{3.16}).  
The second column is based on an exact one-loop evaluation of 
$S_I^{(-)}(s,t)$ and $S_Q^{(\pm)}(s,t)$ from (\ref{2.10}), (\ref{2.11}). 
The third and fourth columns show full one-loop
results and the Born approximation, respectively. 
 }
\label{Table 3}}
\end{table}
\vspace {0.5cm}

\begin{table}[htbp]\centering

{\footnotesize
\begin{tabular}{|r||l|r||l|r||l|l|}\hline
   angle & \multicolumn{2}{c||}{High-energy Born-form}
         & \multicolumn{2}{c||}{Born-form}
         & \multicolumn{1}{c|}{Full one-loop}
         & \multicolumn{1}{c|}{Born} \\ 
 {$(^\circ)$} & \multicolumn{2}{c||}{approximation}
         & \multicolumn{2}{c||}{approximation} & & \\ \hline
         & \multicolumn{1}{c|}{${{d\sigma}\over{d\cos\theta}}$(pb)}
         & \multicolumn{1}{c||}{$\Delta(\%)$}
         & \multicolumn{1}{c|}{${{d\sigma}\over{d\cos\theta}}$(pb)}
         & \multicolumn{1}{c||}{$\Delta(\%)$}
         & \multicolumn{1}{c|}{${{d\sigma}\over{d\cos\theta}}$(pb)}
         & \multicolumn{1}{c|}{${{d\sigma}\over{d\cos\theta}}$(pb)}
            \\  \hline \hline
    & \multicolumn {6}{c|}{$\sigma_{``unpol''}$ } \\ \hline
 10 & 6.166$\times 10^1$     &  +0.04 & 6.164$\times 10^1$    & +0.01 
    & 6.163$\times 10^1$              & 7.502$\times 10^1$ \\ \hline
 30 & 6.475   &$-$0.65 & 6.524   &$-$0.07 & 6.530 & 8.495 \\ \hline
 50 & 1.725   &$-$0.86 & 1.740   &$-$0.25 & 1.746 & 2.432 \\ \hline
 70 & 6.441$\times 10^{-1}$  &$-$0.75 & 6.474$\times 10^{-1}$  &$-$0.41 
    & 6.513$\times 10^{-1}$           & 9.578$\times 10^{-1}$ \\ \hline
 90 & 3.158$\times 10^{-1}$  &$-$0.41 & 3.158$\times 10^{-1}$  &$-$0.41 
    & 3.178$\times 10^{-1}$           & 4.868$\times 10^{-1}$ \\ \hline
110 & 1.869$\times 10^{-1}$  &$-$0.03 & 1.861$\times 10^{-1}$  &$-$0.30 
    & 1.870$\times 10^{-1}$           & 2.974$\times 10^{-1}$ \\ \hline
130 & 1.108$\times 10^{-1}$  &  +0.44 & 1.099$\times 10^{-1}$  &$-$0.05 
    & 1.100$\times 10^{-1}$           & 1.824$\times 10^{-1}$ \\ \hline
150 & 5.599$\times 10^{-2}$  &  +1.56 & 5.507$\times 10^{-2}$ & +0.58 
    & 5.452$\times 10^{-2}$           & 9.418$\times 10^{-2}$ \\ \hline
170 & 2.534$\times 10^{-2}$  &  +3.95 & 2.434$\times 10^{-1}$ &  +1.55
    & 2.369$\times 10^{-2}$           & 4.182$\times 10^{-2}$ \\ \hline
       \hline
    & \multicolumn {6}{c|}{$\sigma_T$ } \\ \hline
 10 & 6.131$\times 10^1$     &  +0.04 & 6.128$\times 10^1$    &  +0.00 
    & 6.128$\times 10^1$              & 7.457$\times 10^1$ \\ \hline
 30 & 6.340   & $-$0.59 & 6.390   &  +0.00 & 6.389 &  8.329 \\ \hline
 50 & 1.585   & $-$0.67 & 1.600   &  +0.00 & 1.600 & 2.250 \\ \hline
 70 & 5.203$\times 10^{-1}$  &$-$0.39 & 5.232$\times 10^{-1}$  &$-$0.02
    & 5.234$\times 10^{-1}$           & 7.915$\times 10^{-1}$ \\ \hline
 90 & 2.153$\times 10^{-1}$  & +0.09  & 2.149$\times 10^{-1}$  &$-$0.03 
    & 2.150$\times 10^{-1}$           & 3.484$\times 10^{-1}$ \\ \hline
110 & 1.117$\times 10^{-1}$  & +0.37  & 1.105$\times 10^{-1}$  &$-$0.26 
    & 1.110$\times 10^{-1}$           & 1.914$\times 10^{-1}$ \\ \hline
130 & 5.960$\times 10^{-2}$  & +0.62  & 5.835$\times 10^{-2}$ &$-$0.54 
    & 5.893$\times 10^{-2}$           & 1.077$\times 10^{-1}$ \\ \hline
150 & 2.350$\times 10^{-2}$  & +1.85  & 2.251$\times 10^{-2}$ &$-$0.54 
    & 2.274$\times 10^{-2}$           & 4.408$\times 10^{-2}$ \\ \hline
170 & 3.486$\times 10^{-3}$  & +17.29 & 2.561$\times 10^{-3}$ &$-$0.21 
    & 2.571$\times 10^{-3}$           & 5.293$\times 10^{-3}$ \\ \hline
    \hline
    & \multicolumn {6}{c|}{$\sigma_L$} \\ \hline
 10 & 1.417$\times 10^{-1}$  & $-$1.02& 1.435$\times 10^{-1}$  & 0.00
    & 1.435$\times 10^{-1}$           & 1.772$\times 10^{-1}$ \\ \hline
 30 & 3.119$\times 10^{-3}$  & $-$5.43& 3.309$\times 10^{-3}$  & 0.00
    & 3.309$\times 10^{-3}$           & 3.497$\times 10^{-3}$ \\ \hline
 50 & 2.620$\times 10^{-2}$  &  +0.73 & 2.599$\times 10^{-2}$  & 0.00
    & 2.599$\times 10^{-2}$           & 2.869$\times 10^{-2}$ \\ \hline
 70 & 5.367$\times 10^{-2}$  & $-$0.12& 5.375$\times 10^{-2}$  & 0.00
    & 5.375$\times 10^{-2}$           & 6.543$\times 10^{-2}$  \\ \hline
 90 & 6.597$\times 10^{-2}$  & $-$0.43& 6.634$\times 10^{-2}$  & 0.00
    & 6.634$\times 10^{-2}$           & 8.630$\times 10^{-2}$  \\ \hline
110 & 5.900$\times 10^{-2}$  & $-$0.44& 5.939$\times 10^{-2}$  & 0.00
    & 5.939$\times 10^{-2}$           & 8.185$\times 10^{-2}$ \\ \hline
130 & 3.847$\times 10^{-2}$  & $-$0.48& 3.866$\times 10^{-2}$  &$-$0.020
    & 3.867$\times 10^{-2}$           & 5.647$\times 10^{-2}$  \\ \hline
150 & 1.571$\times 10^{-2}$  & +0.00  & 1.570$\times 10^{-2}$  &$-$0.04
    & 1.571$\times 10^{-2}$           & 2.454$\times 10^{-2}$ \\ \hline
170 & 1.789$\times 10^{-3}$  & +0.77  & 1.763$\times 10^{-3}$  &$-$0.10
    & 1.766$\times 10^{-3}$           & 2.986$\times 10^{-3}$ \\ 
         \hline \hline
    & \multicolumn {6}{c|} {$\sigma_{mixed}$} \\ \hline
 10 & 2.086$\times 10^{-1}$  &  +1.29 & 2.158$\times 10^{-1}$ &  +3.94 
    & 2.051$\times 10^{-1}$           & 2.714$\times 10^{-1}$ \\ \hline
 30 & 1.318$\times 10^{-1}$  &$-$3.82 & 1.310$\times 10^{-1}$ &$-$4.31 
    & 1.380$\times 10^{-1}$           & 1.625$\times 10^{-1}$ \\ \hline
 50 & 1.142$\times 10^{-1}$  &$-$3.98 & 1.144$\times 10^{-1}$ &$-$3.85 
    & 1.203$\times 10^{-1}$           & 1.531$\times 10^{-1}$ \\ \hline
 70 & 7.017$\times 10^{-2}$  &$-$3.93 & 7.041$\times 10^{-2}$ &$-$3.69 
    & 7.413$\times 10^{-2}$           & 1.008$\times 10^{-1}$ \\ \hline
 90 & 3.459$\times 10^{-2}$  &$-$3.59 & 3.462$\times 10^{-2}$ &$-$3.53 
    & 3.646$\times 10^{-2}$           & 5.216$\times 10^{-2}$  \\ \hline
110 & 1.614$\times 10^{-2}$  &$-$1.86 & 1.616$\times 10^{-2}$ &$-$1.78 
    & 1.659$\times 10^{-2}$           & 2.414$\times 10^{-2}$ \\ \hline
130 & 1.276$\times 10^{-2}$  &  +1.86 & 1.286$\times 10^{-2}$ & +2.41 
    & 1.242$\times 10^{-2}$           & 1.825$\times 10^{-2}$ \\ \hline
150 & 1.678$\times 10^{-2}$  &  +2.78 & 1.686$\times 10^{-2}$ & +3.09 
    & 1.607$\times 10^{-2}$           & 2.556$\times 10^{-2}$ \\ \hline
170 & 2.006$\times 10^{-2}$  &  +2.12 & 2.002$\times 10^{-2}$ & +2.00 
    & 1.935$\times 10^{-2}$           & 3.355$\times 10^{-2}$ \\ \hline
\end{tabular}  
\par
\medskip
\caption{ Same as table 3, but for $\sqrt s = 500$ GeV. } 
\label{ Table 4} }
\end{table}

     A detailed analysis of the differential cross sections at
$\sqrt s = 2000$ GeV and at $\sqrt s = 500$ GeV for various polarization 
states of the $W$-bosons is presented in tables 3 and 4, respectively.
For the $W^{\pm}$-spin-summed, transverse and longitudinal cases, 
the accuracy of the high-energy Born-form approximation at $\sqrt s=2000$ GeV 
is better than $1\%$, except for the production angle of 170$^\circ$, where 
the accuracy is of order 3$\%$.  Even at $\sqrt s=500$ GeV, the accuracy of the 
high-energy Born-form approximation stays below 1$\%$ for most of the angular range. 
For the case of mixed polarizations, because of the strong suppression of the
cross section relative to the sum of purely longitudinal and transverse
production, at $\sqrt s=2000$ GeV, the larger deviations in the 
Born-form approximation are fairly irrelevant, not only for the total 
but also for the differential cross section.  At $\sqrt s = 500$ GeV,
the relative contribution of the mixed polarization is larger 
than in the asymptotic region of $\sqrt s=2000$ GeV, but the accuracy
of the approximation, of order 3$\%$, is also substantially better.\par

     In summary, the high-energy Born-form approximation described in 
the present paper provides, in general, a fully satisfactory description 
of the differential production cross section for various polarization
states of the $W$-bosons at the one-loop level.  We stress again 
the simplicity of the underlying one-loop high-energy 
approximation based on formulae that roughly fill a single page in section 3, 
while the complete expressions need a factor of 10 to 100 more space 
\cite{FJZ}, \cite{BDSBBK}, \cite{K1}. \par
     
\medskip

\section{\bf Conclusion}
    It has been known for some time that a Born-form approximation
with three invariant amplitudes yields a satisfactory
approximation of the differential cross section for W-pair
production at one loop.  
In this paper, we present a novel choice for the one-loop invariant amplitudes 
that is well suited for an approximation in the high-energy limit, including 
W~polarization.
The invariant amplitudes,  $S_I^{(-)}(s,t)$ and 
$S_Q^{(\pm)}(s,t)$, replace the weak and electromagnetic couplings
appearing at tree level, and their high-energy form can be
written down in a few lines.  It is worth stressing that 
a dominant and fairly big contribution to the invariant amplitudes
at high energies is of genuine electroweak origin. This dominant part is 
due to (non-Abelian) 
bosonic loops, but it is fairly insensitive to the value of the Higgs mass, 
as long as the Higgs mass
is constrained to values for which perturbation theory provides a good
approximation.

\vskip 1 truecm

\section*{Acknowledgement}
It is a pleasure to thank Stefan Dittmaier for useful discussions.
\vfill\eject

\renewcommand{\theequation}{\Alph{section}.\arabic{equation}}
\setcounter{section}{1}
\section*{Appendix A. Helicity Amplitudes and High-Energy Behaviour} 
\setcounter{equation}{0}
     In this appendix we briefly recall the origin of the unitarity
constraints (\ref{2.9}).  Furthermore, we give the helicity amplitudes 
in the representations that yield the Born-form amplitudes (\ref{2.10}), 
(\ref{2.11}) and (\ref{2.13}), (\ref{2.14}), respectively.\par
     The unitarity constraints (\ref{2.9}) on the invariant amplitudes follow
from inspection of table A1, which shows the high-energy approximation
of the basic matrix elements, ${\cal M}_I^{(\sigma)}$, 
${\cal M}_Q^{(\sigma)}$ and $\bar M_i^{(\sigma)} (i=2,3,4,6)$.  
The requirement of a unitarity-preserving high-energy behaviour of the 
helicity amplitudes, together with the linear independence of the 
column vectors of the matrix in table A1, implies the high-energy 
constraints (\ref{2.9}).\par

\vskip 1 truecm
\begin{tabular}{|l|l|l|l|l|}\hline
           &     &      &      &      \\
  ($\sigma;\lambda,\bar\lambda$)& ${\cal M}_I$ & ${\cal M}_Q$ & 
          ${1\over s}\bar M_2$ & ${1\over s}\bar M_3$  \\ 
           &     &      &      &      \\ \hline
  ($\pm$;00) & $ {{1-2c_W^2}\over{\sqrt 2 c_W^2}}$ & 
               $- {1\over{\sqrt 2c_W^2}}$  &  
               $-{s\over{\sqrt 2 M_W^2}}$  &  0  \\ \hline
  ($\pm;--$)($\pm;++$) & $\sqrt 2{{M_W^2}\over s}
              [{{1-4c_W^2}\over{c_W^2}}+{2\over{1-\cos\theta}}]$  &
               $-\sqrt 2{{M_Z^2}\over s}$ &  0 & 
               ${s\over{\sqrt 2 M_W^2}}$  \\ \hline
  ($\pm;\mp \pm$) & ${2\over{1-\cos\theta}}$ & 0 & 0 & 0 \\ \hline  
  ($\pm;\pm \mp$) & ${2\over{1-\cos\theta}}$ & 0 & 0 & 0 \\ \hline
  ($+;-0$)($+;0+$)& ${{2M_W}\over{\sqrt{2s}}}
                    [{{1-4c_W^2}\over{c_W^2}}+{4\over{1-\cos\theta}}]$ &
                    $-{{2M_W}\over{\sqrt{2s}c_W^2}}$ & 
                    $-{{\sqrt s}\over{\sqrt 2 M_W}}$ &
                    ${{\sqrt s}\over{\sqrt 2 M_W}}$   \\ 
  ($-:0-$)($-;+0$)&   &   &   &  \\ \hline
  ($+;0-$)($+;+0$)& ${{2M_W}\over{\sqrt{2s}}}[{1\over{c_W^2}}-4]$ & 
                    $-{{2M_W}\over{\sqrt{2s}c_W^2}}$ &  
                    $-{{\sqrt s}\over{\sqrt 2 M_W}}$ &
                    ${{\sqrt s}\over{\sqrt 2 M_W}}$  \\ 
  ($-:-0$)($-;0+$)&   &   &   &  \\ \hline
\end{tabular}

\medskip

\begin{tabular}{|l|l|l|l|}\hline
        &   &   &     \\
  ($\sigma;\lambda,\bar\lambda$) & ${1\over s}\bar M_4$ &
    ${1\over s}\bar M_6$ & $d^{J_0}_{\Delta\sigma,\Delta \lambda}$  \\ 
        &   &   &     \\ \hline
  ($\pm$;00) &  0 & $- {s\over{M_W^2}}~{{\cos^2\theta}\over{2\sqrt 2}}$ &
                    $\mp{{\sin\theta}\over{\sqrt 2}}$  \\ \hline
  ($\pm;--$)($\pm;++$) & 0 & ${{\sin^2\theta}\over {\sqrt 2}}$ &
                             $\mp{{\sin\theta}\over{\sqrt 2}}$ \\ \hline
  ($\pm;\mp \pm$) & 0 & $-(1+\cos\theta)$ & 
                    $\mp(1-\cos\theta){{\sin\theta}\over 2}$  \\ \hline
  ($\pm;\pm \mp$) & 0 & $1-\cos\theta$ &
                     $\pm(1+\cos\theta){{\sin\theta}\over 2}$ \\ \hline
  ($+;-0$)($+;0+$) & 
      $-\sigma{{\sqrt s}\over{\sqrt 2 M_W}}{s\over{M_W^2}}$ &
      $-{{\sqrt s}\over{\sqrt 2 M_W}}(1+\cos\theta)\cos\theta$ & 
                          ${1\over 2}(1-\cos\theta)$ \\ 
  ($-:0-$)($-;+0$) &   &  &  \\ \hline
  ($+;0-$)($+;+0$) & 
      $\sigma{{\sqrt s}\over{\sqrt 2 M_W}}{s\over{M_W^2}}$  &
      ${{\sqrt s}\over{\sqrt 2 M_W}}(1-\cos\theta)\cos\theta$ & 
                          ${1\over 2}(1+\cos\theta)$  \\
  ($-:-0$)($-;0+$) &  &  &  \\ \hline
\end{tabular}
\par
\medskip

{\footnotesize
\noindent{\bf Table A1}.  The high-energy behaviour of ${\cal M}_I$, 
${\cal M}_Q$ and ${1\over s}\bar M_i$, which is obtained by multiplying
each entry by the factor $d^{J_0}_{\Delta\sigma,\Delta\lambda}$ given
in the rightmost column.}

\vfill\eject

     We turn to the helicity amplitudes in a representation that contains
the Born form (\ref{2.8}) with  $S_I^{(\sigma)}(s,t)$ and 
$S_Q^{(\sigma)}(s,t)$ from (\ref{2.10}), (\ref{2.11}) as an approximation.
The helicity amplitudes (\ref{2.5}) may in fact be rewritten identically 
in the form
\bq
    {\cal H}(\sigma, \lambda,\bar\lambda) =
          S_I^{(\sigma)}(s,t){\cal M}_I(\sigma,\lambda,\bar\lambda) 
         +S_Q^{(\sigma)}(s,t){\cal M}_Q(\sigma,\lambda,\bar\lambda)
         +\sum_{i=3}^6S_i^{(\sigma)}(s,t) 
           {\cal M}_i(\sigma,\lambda,\bar\lambda) , \label{A.1}
\eq
with  $S_I^{(\sigma)}(s,t)$ and $S_Q^{(\sigma)}(s,t)$ in the form 
(\ref{2.10}), (\ref{2.11}), i.e.
\bqa
    S_I^{(\sigma)}&=& -tY^{(\sigma)}_5, \label{A.2} \\
    S_Q^{(\sigma)}&=& -{{st}\over{M_Z^2}}Y^{(\sigma)}_5 
         +{{s(s-M_Z^2)}\over{M_Z^2}}  [Y^{(\sigma)}_1
         + {2\over{3-\beta^2}}Y^{(\sigma)}_2 
         + {{\cos^2\theta}\over{3-\beta^2}}Y^{(\sigma)}_6], \label{A.3} 
\eqa
and
\bqa
    S_3^{(\sigma)}&=&s Y^{(\sigma)}_2, \label{A.4}\\
    S_4^{(\sigma)}&=&s Y^{(\sigma)}_3, \label{A.5}\\
    S_5^{(\sigma)}&=&s Y^{(\sigma)}_4, \label{A.6}\\ 
    S_6^{(\sigma)}&=&s Y^{(\sigma)}_6, \label{A.7}
\eqa
with
\bqa
    {\cal M}_3 &=& {1\over s} \bar M_2 - {{s-M_Z^2}\over {M_Z^2}}~
                 {2\over{3-\beta^2}}{\cal M}_Q
               =  {1\over s} \bar M_2 - {1\over s}~ 
                 {2\over{3-\beta^2}}\bar M_1, \label{A.8} \\
    {\cal M}_4 &=& {1\over s} \bar M_3, \label{A.9} \\              
    {\cal M}_5 &=& {1\over s} \bar M_4, \label{A.10} \\              
    {\cal M}_6 &=& {1\over s} \bar M_6 - {{s-M_Z^2}\over {M_Z^2}}~
                 {{\cos^2\theta}\over{3-\beta^2}}{\cal M}_Q
               =  {1\over s} \bar M_6 - {1\over s}~ 
                 {{\cos^2\theta}\over{3-\beta^2}}\bar M_1. \label{A.11}
\eqa
In this basis, ${\cal M}_i(\sigma,\lambda=\bar\lambda=0)=0$ for $i$ = 3, 4, 5, 6;
the longitudinal helicity amplitudes reduce to the Born form,
except for the presence of $S^{(+)}_I$ which at one loop turns out to
be suppressed by several orders of magnitude relative to the other
Born-form invariant amplitudes.
\par
     We finally give the helicity amplitudes in the form that contains the 
     Born-form approximation from refs. \cite{FK3S}, \cite{DBD} given by 
(\ref{2.13}), (\ref{2.14}):
\bqa
    S_I^{(\sigma)}&=& -tY^{(\sigma)}_5, \label{A.12} \\
    S_Q^{(\sigma)} &=& -{{ts}\over{M_Z^2}}Y_5^{(\sigma)}
                   +{{s(s-M_Z^2)}\over{M_Z^2}}
                   [Y_1^{(\sigma)} + {1\over 2}Y_2^{(\sigma)}
                    -{1\over 2}Y_3^{(\sigma)}
                    -\sigma{1\over 2}{s\over{M_W^2}} \cos\theta Y_4^{(\sigma)}]
                    \nonumber \\
               &=& {{g^2}\over 2} {s\over{M_Z^2}}
                  (\delta_{\sigma,-} + \delta_t^{(\sigma)}) \nonumber \\
               & & +{{s(s-M_Z^2)}\over{M_Z^2}}[ -e^2{1\over s}(1+\delta_\gamma
                     +{1\over 2}(x_\gamma-y_\gamma
                     -\sigma \cos\theta{s\over{M_W^2}}z_\gamma)) 
                      \nonumber \\
               & & +(e^2-{{g^2}\over 2}\delta_{\sigma,-})
                   {1\over {s-M_Z^2}} \{1+{{s_W}\over{c_W}}(\delta_Z+
                   {1\over 2}(x_Z-y_Z
                - \sigma \cos\theta{s\over{M_W^2}}z_Z))\} ], \label{A.13} 
\eqa
with $S_i^{(\sigma)}(i=3,4,5,6)$ from (\ref{A.4})--(\ref{A.7})  and
\bqa
 {\cal M}_3 &=& {1\over s}\bar M_2 -{{s-M_Z^2}\over{2M_Z^2}} {\cal M}_Q
             =  {1\over s}\bar M_2 -{1\over {2s}}\bar M_1, \label{A.14}\\
 {\cal M}_4 &=& {1\over s}\bar M_3 +{{s-M_Z^2}\over{2M_Z^2}} {\cal M}_Q
             =  {1\over s}\bar M_3 +{1\over {2s}}\bar M_1, \label{A.15}\\
 {\cal M}_5 &=& {1\over s}\bar M_4 +{{s-M_Z^2}\over{2M_Z^2}} 
               \sigma\cos\theta{s\over{M_W^2}}{\cal M}_Q
             =  {1\over s}\bar M_4 +{1\over{2M_W^2}}
               \sigma\cos\theta\bar M_1, \label{A.16}\\
 {\cal M}_6 &=& {1\over s}\bar M_6. \label{A.17}
\eqa

\vskip 1 truecm

\section*{Appendix B. Numerical Results in Different Born-Form Approximations}
     In this appendix we compare the numerical one-loop results for cross sections 
in the different  Born-form approximations based on 
(\ref{2.10}), (\ref{2.11}) and on (\ref{2.13}), (\ref{2.14}).
The numerical results were obtained by employing the computer code from
ref.~\cite{K1}. Soft-photon bremsstrahlung is included as described in
sections 3 to 5.
The purpose of this investigation is twofold:\par
i) to show that the form (\ref{2.10}), (\ref{2.11}) for $S_I^{(-)}(s,t)$ and
$S_Q^{(\pm)}(s,t)$ yields particularly good results in the high-energy 
limit, and\par
ii) to establish the connection with previous work, based on the form 
(\ref{2.13}), (\ref{2.14}).\par
The results for the differential cross sections at $\sqrt s=2000$ GeV,
$\sqrt s=500$ GeV and
$\sqrt s = 200$ GeV are shown in tables B1, B2 and B3.  One observes that 
the cross section with summation over $W^+$ and $W^-$ spins and the 
cross section for production of purely longitudinal $W$-bosons 
in the high-energy limit ($\sqrt s \cong 2000$ GeV) are better
approximated in the Born-form approximation of the present paper, based on
$S_I^{(-)}(s,t)$ and $S_Q^{(\pm)}(s,t)$ from (\ref{2.10}), (\ref{2.11}).

\newpage
                           
{\footnotesize
\begin{tabular}{|r||l|r||l|r||l|l|}\hline
   Angle & \multicolumn{2}{c||}{Born-form approx.}
         & \multicolumn{2}{c||}{Born-form approx. (\cite
{FK3S},\cite{DBD})}
         & \multicolumn{1}{c|}{Full one-loop}
         & \multicolumn{1}{c|}{Born} \\ \hline
{$(^\circ)$} & \multicolumn{1}{c|}{${{d\sigma}\over{d\cos\theta}}$(pb)}
         & \multicolumn{1}{c||}{$\Delta(\%)$}
         & \multicolumn{1}{c|}{${{d\sigma}\over{d\cos\theta}}$(pb)}
         & \multicolumn{1}{c||}{$\Delta(\%)$}
         & \multicolumn{1}{c|}{${{d\sigma}\over{d\cos\theta}}$(pb)}
         & \multicolumn{1}{c|}{${{d\sigma}\over{d\cos\theta}}$(pb)}
            \\  \hline \hline
     & \multicolumn {6}{c|}{$\sigma_{``unpol}$''} \\ \hline
  10 & 3.751    &  +0.02& 3.752   &  +0.04 & 3.750     & 5.329 \\ \hline 
  30 & 2.599$\times 10^{-1}$    &  +0.26& 2.604$\times 10^{-1}$   &  +0.36 
     & 2.586$\times 10^{-1}$            & 5.070$\times 10^{-1}$ \\ \hline 
  50 & 5.551$\times 10^{-2}$    &  +0.37& 5.602$\times 10^{-2}$  &  +0.74 
     & 5.500$\times 10^{-2}$            & 1.382$\times 10^{-1}$ \\ \hline 
  70 & 1.727$\times 10^{-2}$    &  +0.21& 1.758$\times 10^{-2}$  &  +0.81 
     & 1.716$\times 10^{-2}$            & 5.208$\times 10^{-2}$ \\ \hline 
  90 & 7.565$\times 10^{-3}$    &$-$0.10& 7.597$\times 10^{-3}$  &  +0.02 
     & 7.592$\times 10^{-3}$            & 2.605$\times 10^{-2}$ \\ \hline 
 110 & 4.074$\times 10^{-3}$    &$-$0.38& 3.898$\times 10^{-3}$  &$-$1.48 
     & 4.135$\times 10^{-3}$            & 1.599$\times 10^{-2}$ \\ \hline 
 130 & 2.007$\times 10^{-3}$    &$-$0.49& 1.788$\times 10^{-3}$  &$-$2.80 
     & 2.053$\times 10^{-3}$            & 9.466$\times 10^{-3}$ \\ \hline 
 150 & 6.589$\times 10^{-4}$    &$-$0.22& 5.319$\times 10^{-4}$  &$-$3.40 
     & 6.678$\times 10^{-4}$            & 3.993$\times 10^{-3}$ \\ \hline 
 170 & 7.567$\times 10^{-5}$    &  +1.76& 5.133$\times 10^{-5}$&$-$2.33 
     & 6.520$\times 10^{-5}$            & 5.961$\times 10^{-4}$ \\ \hline 
       \hline
      & \multicolumn {6}{c|}{$\sigma_T $ } \\ \hline
   10 & 3.751     &  +0.04 & 3.751     &  +0.04 & 3.749     & 5.329 \\ \hline
   30 & 2.585$\times 10^{-1}$  & +0.30  & 2.585$\times 10^{-1}$  & +0.30 
      & 2.570$\times 10^{-1}$           & 5.050$\times 10^{-1}$ \\ \hline
   50 & 5.330$\times 10^{-2}$  & +0.45  & 5.330$\times 10^{-2}$  &  +0.45 
      & 5.270$\times 10^{-2}$           & 1.347$\times 10^{-1}$ \\ \hline
   70 & 1.454$\times 10^{-2}$  & +0.36  & 1.454$\times 10^{-2}$  & +0.36 
      & 1.437$\times 10^{-2}$           & 4.720$\times 10^{-2}$ \\ \hline
   90 & 4.875$\times 10^{-3}$  &$-$0.00 & 4.875$\times 10^{-3}$  &$-$0.00 
      & 4.876$\times 10^{-3}$           & 2.072$\times 10^{-2}$ \\ \hline
  110 & 1.956$\times 10^{-3}$  &$-$0.48 & 1.956$\times 10^{-3}$  &$-$0.48 
      & 2.010$\times 10^{-3}$           & 1.135$\times 10^{-2}$ \\ \hline
  130 & 7.400$\times 10^{-4}$  &$-$0.80 & 7.400$\times 10^{-4}$  &$-$0.80 
      & 7.907$\times 10^{-4}$           & 6.367$\times 10^{-3}$ \\ \hline
  150 & 1.689$\times 10^{-4}$  &$-$0.76 & 1.689$\times 10^{-4}$  &$-$0.76 
      & 1.887$\times 10^{-4}$           & 2.603$\times 10^{-3}$\\ \hline
  170 & 5.448$\times 10^{-6}$  &$-$0.26 & 5.447$\times 10^{-6}$ & $-$0.26 
      & 6.262$\times 10^{-6}$           & 3.124$\times 10^{-4}$ \\
          \hline \hline
      & \multicolumn {6}{c|}{$\sigma_L $} \\ \hline
   10 & 4.329$\times 10^{-5}$  &   0.00 & 9.403$\times 10^{-5}$ &+124.24 
      & 4.329$\times 10^{-5}$           & 4.084$\times 10^{-5}$ \\ \hline
   30 & 8.894$\times 10^{-4}$  &   0.00 & 1.315$\times 10^{-3}$  & +37.14 
      & 8.894$\times 10^{-4}$           & 1.146$\times 10^{-3}$ \\ \hline
   50 & 1.915$\times 10^{-3}$  & 0.00   & 2.395$\times 10^{-3}$ & +16.46 
      & 1.915$\times 10^{-3}$           & 2.917$\times 10^{-3}$ \\ \hline
   70 & 2.580$\times 10^{-3}$  &   0.00 & 2.883$\times 10^{-3}$ & +6.75 
      & 2.580$\times 10^{-3}$           & 4.487$\times 10^{-3}$ \\ \hline
   90 & 2.625$\times 10^{-3}$  &   0.00 & 2.656$\times 10^{-3}$  & +0.60 
      & 2.625$\times 10^{-3}$           & 5.126$\times 10^{-3}$  \\ \hline
  110 & 2.085$\times 10^{-3}$  &   0.00 & 1.911$\times 10^{-3}$  & $-$3.83 
      & 2.085$\times 10^{-3}$           & 4.546$\times 10^{-3}$ \\ \hline
  130 & 1.236$\times 10^{-3}$  &   0.00 & 1.021$\times 10^{-3}$  & $-$7.10 
      & 1.236$\times 10^{-3}$           & 3.028$\times 10^{-3}$ \\ \hline
  150 & 4.528$\times 10^{-4}$  &  +0.01 & 3.349$\times 10^{-4}$  & $-$9.12 
      & 4.527$\times 10^{-4}$           & 1.292$\times 10^{-4}$ \\ \hline
  170 & 3.834$\times 10^{-5}$  & $-$0.04& 2.499$\times 10^{-5}$ & $-$8.60 
      & 3.840$\times 10^{-5}$           & 1.559$\times 10^{-4}$ \\ 
           \hline \hline
    & \multicolumn {6}{c|} {$\sigma_{mixed}$} \\ \hline
   10 & 6.425$\times 10^{-4}$  &$-$39.22 & 9.313$\times 10^{-4}$ &$-$4.15 
      & 9.655$\times 10^{-4}$            & 8.236$\times 10^{-4}$ \\ \hline
   30 & 4.946$\times 10^{-4}$  &$-$19.68 & 5.968$\times 10^{-4}$ &$-$7.58 
      & 6.608$\times 10^{-4}$            & 8.444$\times 10^{-4}$ \\ \hline
   50 & 2.959$\times 10^{-4}$  &$-$14.97 & 3.301$\times 10^{-4}$ &$-$9.58 
      & 3.907$\times 10^{-4}$            & 6.332$\times 10^{-4}$ \\ \hline
   70 & 1.500$\times 10^{-4}$  &$-$13.65 & 1.583$\times 10^{-4}$ &$-$11.54
      & 2.037$\times 10^{-4}$            & 3.933$\times 10^{-4}$ \\ \hline
   90 & 6.528$\times 10^{-5}$  &$-$13.12 & 6.567$\times 10^{-5}$ &$-$12.92
      & 9.151$\times 10^{-5}$            & 1.999$\times 10^{-4}$  \\ \hline
  110 & 3.252$\times 10^{-5}$  &$-$10.57 & 3.088$\times 10^{-5}$ &$-$9.91 
      & 4.010$\times 10^{-5}$            & 9.307$\times 10^{-5}$ \\ \hline
  130 & 3.082$\times 10^{-5}$  &   +6.10 & 2.637$\times 10^{-5}$ &$-$0.17 
      & 2.649$\times 10^{-5}$            & 7.099$\times 10^{-5}$ \\ \hline
  150 & 3.723$\times 10^{-5}$  &  +11.05 & 2.808$\times 10^{-5}$ &  +1.73
      & 2.638$\times 10^{-5}$            & 9.818$\times 10^{-5}$ \\ \hline
  170 & 3.188$\times 10^{-5}$  &   +8.87 & 2.089$\times 10^{-5}$ &  +0.27 
      & 2.054$\times 10^{-5}$            & 1.278$\times 10^{-4}$ \\ \hline
\end{tabular}
}\par
\medskip
{\footnotesize
\noindent{\bf Table B1}. The differential cross section 
${{d\sigma}\over{d\cos\theta}}$ and the deviations from the 
full one-loop result  are shown at $\sqrt s = 2000$ GeV,
when summing over $W$ polarizations, for 
transversely polarized $W$,
longitudinally polarized $W$,  and the mixed case. 
The first column shows the result of
the Born-form approximation based on the one-loop evaluation of 
$S_I^{(-)}$ and $S_Q^{(\pm)}$ 
given by (\ref{2.10}) and (\ref{2.11}).  
The second column gives the result of previous
work \cite{FK3S}, \cite{DBD}, 
which is based on the one-loop evaluation of
$S_I^{(-)}$ and $S_Q^{(\pm)}$ given by (\ref{2.13}), (\ref{2.14}).
The third and fourth columns give the full one-loop results and the
Born approximation.
}

\vfill\eject

{\footnotesize
\begin{tabular}{|r||l|r||l|r||l|l|}\hline
   Angle & \multicolumn{2}{c||}{Born-form approx.}
         & \multicolumn{2}{c||}{Born-form approx. (\cite{FK3S},\cite{DBD})}
         & \multicolumn{1}{c|}{Full one-loop}
         & \multicolumn{1}{c|}{Born} \\ \hline
{$(^\circ)$} & \multicolumn{1}{c|}{${{d\sigma}\over{d\cos\theta}}$(pb)}
         & \multicolumn{1}{c||}{$\Delta(\%)$}
         & \multicolumn{1}{c|}{${{d\sigma}\over{d\cos\theta}}$(pb)}
         & \multicolumn{1}{c||}{$\Delta(\%)$}
         & \multicolumn{1}{c|}{${{d\sigma}\over{d\cos\theta}}$(pb)}
         & \multicolumn{1}{c|}{${{d\sigma}\over{d\cos\theta}}$(pb)}
            \\  \hline \hline
    & \multicolumn{6}{c|}{$\sigma_{``unpol''}$} \\ \hline
   10 & 6.164$\times 10^1$     &  +0.01 & 6.163$\times 10^1$      & +0.00 
      & 6.163$\times 10^1$              & 7.502$\times 10^1$ \\ \hline
   30 & 6.524   &$-$0.07 & 6.529   &$-$0.01 & 6.530 & 8.495 \\ \hline
   50 & 1.740   &$-$0.25 & 1.745   &$-$0.04 & 1.746 & 2.432 \\ \hline
   70 & 6.474$\times 10^{-1}$  &$-$0.41 & 6.509$\times 10^{-1}$  &$-$0.04 
      & 6.513$\times 10^{-1}$           & 9.578$\times 10^{-1}$ \\ \hline
   90 & 3.158$\times 10^{-1}$  &$-$0.41 & 3.169$\times 10^{-1}$  &$-$0.18 
      & 3.178$\times 10^{-1}$           & 4.868$\times 10^{-1}$ \\ \hline
  110 & 1.861$\times 10^{-1}$  &$-$0.30 & 1.852$\times 10^{-1}$  &$-$0.61 
      & 1.870$\times 10^{-1}$           & 2.974$\times 10^{-1}$ \\ \hline
  130 & 1.099$\times 10^{-1}$  &$-$0.05 & 1.083$\times 10^{-1}$  &$-$0.93 
      & 1.100$\times 10^{-1}$           & 1.824$\times 10^{-1}$ \\ \hline
  150 & 5.507$\times 10^{-2}$  &  +0.58 & 5.374$\times 10^{-2}$ &$-$0.83 
      & 5.452$\times 10^{-2}$           & 9.418$\times 10^{-2}$ \\ \hline
  170 & 2.434$\times 10^{-2}$  &  +1.55 & 2.359$\times 10^{-2}$ &$-$0.24 
      & 2.369$\times 10^{-2}$           & 4.182$\times 10^{-2}$ \\ \hline
       \hline
    & \multicolumn {6}{c|}{$\sigma_T$ } \\ \hline
   10 & 6.128$\times 10^1$     &  +0.00 & 6.128$\times 10^1$    &  +0.00 
      & 6.128$\times 10^1$              & 7.457$\times 10^1$ \\ \hline
   30 & 6.390   &  +0.00 & 6.389   &  +0.00 & 6.389 &  8.329 \\ \hline
   50 & 1.600   &  +0.00 & 1.600   &  +0.00 & 1.600 & 2.250 \\ \hline
   70 & 5.232$\times 10^{-1}$  &$-$0.02 & 5.232$\times 10^{-1}$  &$-$0.02 
      & 5.234$\times 10^{-1}$           & 7.915$\times 10^{-1}$ \\ \hline
   90 & 2.149$\times 10^{-1}$  &$-$0.03 & 2.149$\times 10^{-1}$  &$-$0.03 
      & 2.150$\times 10^{-1}$           & 3.484$\times 10^{-1}$ \\ \hline
  110 & 1.105$\times 10^{-1}$  &$-$0.26 & 1.105$\times 10^{-1}$  &$-$0.26 
      & 1.110$\times 10^{-1}$           & 1.914$\times 10^{-1}$ \\ \hline
  130 & 5.835$\times 10^{-2}$  &$-$0.54 & 5.834$\times 10^{-2}$ &$-$0.54 
      & 5.893$\times 10^{-2}$           & 1.077$\times 10^{-1}$ \\ \hline
  150 & 2.251$\times 10^{-2}$  &$-$0.54 & 2.250$\times 10^{-2}$ &$-$0.54 
      & 2.274$\times 10^{-2}$           & 4.408$\times 10^{-2}$ \\ \hline
  170 & 2.561$\times 10^{-3}$  &$-$0.21 & 2.560$\times 10^{-3}$ &$-$0.21 
      & 2.571$\times 10^{-3}$           & 5.293$\times 10^{-3}$ \\ \hline
    \hline
    & \multicolumn {6}{c|}{$\sigma_L$} \\ \hline
   10 & 1.435$\times 10^{-1}$  &   0.00 & 1.407$\times 10^{-1}$  &$-$1.58 
      & 1.435$\times 10^{-1}$           & 1.772$\times 10^{-1}$ \\ \hline
   30 & 3.309$\times 10^{-3}$  &   0.00 & 3.098$\times 10^{-3}$  &$-$6.03 
      & 3.309$\times 10^{-3}$           & 3.497$\times 10^{-3}$ \\ \hline
   50 & 2.599$\times 10^{-2}$  &   0.00 & 2.823$\times 10^{-2}$  &  +7.81 
      & 2.599$\times 10^{-2}$           & 2.869$\times 10^{-2}$ \\ \hline
   70 & 5.375$\times 10^{-2}$  &   0.00 & 5.619$\times 10^{-2}$  &  +3.73 
      & 5.375$\times 10^{-2}$           & 6.543$\times 10^{-2}$  \\ \hline
   90 & 6.634$\times 10^{-2}$  &   0.00 & 6.728$\times 10^{-2}$  &  +1.09 
      & 6.634$\times 10^{-2}$           & 8.630$\times 10^{-2}$  \\ \hline
  110 & 5.939$\times 10^{-2}$  &   0.00 & 5.872$\times 10^{-2}$  &$-$0.82 
      & 5.939$\times 10^{-2}$           & 8.185$\times 10^{-2}$ \\ \hline
  130 & 3.866$\times 10^{-2}$  & $-$0.02& 3.748$\times 10^{-2}$  &$-$2.11 
      & 3.867$\times 10^{-2}$           & 5.647$\times 10^{-2}$  \\ \hline
  150 & 1.570$\times 10^{-2}$  & $-$0.04& 1.509$\times 10^{-2}$  &$-$2.53 
      & 1.571$\times 10^{-2}$           & 2.454$\times 10^{-2}$ \\ \hline
  170 & 1.763$\times 10^{-3}$  & $-$0.01& 1.704$\times 10^{-3}$  &$-$2.08 
      & 1.766$\times 10^{-3}$           & 2.986$\times 10^{-3}$ \\ 
         \hline \hline
    & \multicolumn {6}{c|} {$\sigma_{mixed}$} \\ \hline
   10 & 2.158$\times 10^{-1}$  &  +3.94 & 2.052$\times 10^{-1}$ &  +0.04 
      & 2.051$\times 10^{-1}$           & 2.714$\times 10^{-1}$ \\ \hline
   30 & 1.310$\times 10^{-1}$  &$-$4.31 & 1.363$\times 10^{-1}$ &$-$1.05 
      & 1.380$\times 10^{-1}$           & 1.625$\times 10^{-1}$ \\ \hline
   50 & 1.144$\times 10^{-1}$  &$-$3.85 & 1.175$\times 10^{-1}$ &$-$1.83 
      & 1.203$\times 10^{-1}$           & 1.531$\times 10^{-1}$ \\ \hline
   70 & 7.041$\times 10^{-2}$  &$-$3.69 & 7.147$\times 10^{-2}$ &$-$2.64 
      & 7.413$\times 10^{-2}$           & 1.008$\times 10^{-1}$ \\ \hline
   90 & 3.462$\times 10^{-2}$  &$-$3.53 & 3.480$\times 10^{-2}$ &$-$3.18 
      & 3.646$\times 10^{-2}$           & 5.216$\times 10^{-2}$  \\ \hline
  110 & 1.616$\times 10^{-2}$  &$-$1.78 & 1.604$\times 10^{-2}$ &$-$2.28 
      & 1.659$\times 10^{-2}$           & 2.414$\times 10^{-2}$ \\ \hline
  130 & 1.286$\times 10^{-2}$  &  +2.41 & 1.246$\times 10^{-2}$ & +0.22 
      & 1.242$\times 10^{-2}$           & 1.825$\times 10^{-2}$ \\ \hline
  150 & 1.686$\times 10^{-2}$  &  +3.09 & 1.615$\times 10^{-2}$ & +0.31 
      & 1.607$\times 10^{-2}$           & 2.556$\times 10^{-2}$ \\ \hline
  170 & 2.002$\times 10^{-2}$  &  +2.00 & 1.932$\times 10^{-2}$ &$-$0.09 
      & 1.935$\times 10^{-2}$           & 3.355$\times 10^{-2}$ \\ \hline
\end{tabular}
}\par
\medskip
{\footnotesize
\noindent{\bf Table B2}. Same as table B1, but for $\sqrt s = 500$ GeV.}

\newpage

{\footnotesize
\begin{tabular}{|r||l|r||l|r||l|l|}\hline
   Angle & \multicolumn{2}{c||}{Born-form approx.}
         & \multicolumn{2}{c||}{Born-form approx. (\cite{FK3S},\cite{DBD})}
         & \multicolumn{1}{c|}{Full one-loop}
         & \multicolumn{1}{c|}{Born} \\ \hline
{$(^\circ)$} & \multicolumn{1}{c|}{${{d\sigma}\over{d\cos\theta}}$(pb)}
         & \multicolumn{1}{c||}{$\Delta(\%)$}
         & \multicolumn{1}{c|}{${{d\sigma}\over{d\cos\theta}}$(pb)}
         & \multicolumn{1}{c||}{$\Delta(\%)$}
         & \multicolumn{1}{c|}{${{d\sigma}\over{d\cos\theta}}$(pb)}
         & \multicolumn{1}{c|}{${{d\sigma}\over{d\cos\theta}}$(pb)}
            \\  \hline \hline
    & \multicolumn{6}{c|}{$\sigma_{``unpol''}$} \\ \hline
   10 & 3.408$\times 10^1$     &$-$0.12 & 3.413$\times 10^1$   & 0.00 
      & 3.413$\times 10^1$              & 4.141$\times 10^1$ \\ \hline
   30 & 2.310$\times 10^1$     &$-$0.04 & 2.311$\times 10^1$   & 0.00 
      & 2.311$\times 10^1$              & 2.813$\times 10^1$ \\ \hline
   50 & 1.255$\times 10^1$     &$-$0.06 & 1.255$\times 10^1$   &$-$0.06 
      & 1.256$\times 10^1$              & 1.540$\times 10^1$ \\ \hline
   70 & 6.958                  &$-$0.05 & 6.960                &$-$0.02 
      & 6.962                           & 8.617              \\ \hline
   90 & 4.197                  &$-$0.04 & 4.200                &  +0.02
      & 4.199                           & 5.251              \\ \hline
  110 & 2.766                  &$-$0.06 & 2.768                &   0.00 
      & 2.768                           & 3.494              \\ \hline
  130 & 1.964                  &$-$0.08 & 1.965                &$-$0.04 
      & 1.966                           & 2.502              \\ \hline
  150 & 1.502                  &$-$0.16 & 1.503                &$-$0.10 
      & 1.505                           & 1.927              \\ \hline
  170 & 1.283                  &$-$0.18 & 1.284                &$-$0.10 
      & 1.286                           & 1.652               \\ \hline
       \hline
    & \multicolumn {6}{c|}{$\sigma_T$ } \\ \hline
   10 & 7.998                  &  +0.00 & 7.999                 &  +0.01 
      & 7.998                           & 9.644              \\ \hline
   30 & 1.828$\times 10^1$     &  +0.00 & 1.828$\times 10^1$    &  +0.00 
      & 1.828$\times 10^1$              & 2.218$\times 10^1$ \\ \hline
   50 & 9.977                  &$-$0.08 & 9.977                 &$-$0.08 
      & 9.987                           & 1.225$\times 10^1$ \\ \hline
   70 & 4.364                  &$-$0.13 & 4.364                 &$-$0.13 
      & 4.371                           & 5.432               \\ \hline
   90 & 2.166                  &  +0.00 & 2.166                 &  +0.00 
      & 2.166                           & 2.735               \\ \hline
  110 & 1.349                  &  +0.06 & 1.349                 &  +0.06 
      & 1.348                           & 1.728               \\ \hline
  130 & 8.329$\times 10^{-1}$  &$-$0.04 & 8.330$\times 10^{-1}$ &$-$0.03 
      & 8.333$\times 10^{-1}$           & 1.083               \\ \hline
  150 & 3.576$\times 10^{-1}$  &$-$0.11 & 3.576$\times 10^{-1}$ &$-$0.11 
      & 3.581$\times 10^{-1}$           & 4.710$\times 10^{-1}$ \\ \hline
  170 & 4.370$\times 10^{-2}$  &$-$0.14 & 4.370$\times 10^{-2}$ &$-$0.14 
      & 4.378$\times 10^{-2}$           & 5.801$\times 10^{-2}$ \\ \hline
    \hline
    & \multicolumn {6}{c|}{$\sigma_L$} \\ \hline
   10 & 1.377                  &  +0.00 & 1.379                  &  +0.12
      & 1.377                           & 1.671                 \\ \hline
   30 & 2.325                  &  +0.00 & 2.328                  &  +0.11 
      & 2.325                           & 2.856                 \\ \hline
   50 & 4.886$\times 10^{-1}$  &  +0.06 & 4.880$\times 10^{-1}$  &$-$0.03 
      & 4.882$\times 10^{-1}$           & 6.176$\times 10^{-1}$ \\ \hline
   70 & 4.896$\times 10^{-2}$  &  +0.20 & 4.910$\times 10^{-2}$  &  +0.47 
      & 4.886$\times 10^{-2}$           & 5.087$\times 10^{-2}$  \\ \hline
   90 & 3.791$\times 10^{-1}$  & $-$0.09& 3.807$\times 10^{-1}$  &  +0.27 
      & 3.795$\times 10^{-1}$           & 4.495$\times 10^{-1}$  \\ \hline
  110 & 6.471$\times 10^{-1}$  & $-$0.11& 6.483$\times 10^{-1}$  &  +0.04 
      & 6.480$\times 10^{-1}$           & 7.971$\times 10^{-1}$ \\ \hline
  130 & 5.735$\times 10^{-1}$  & $-$0.12& 5.739$\times 10^{-1}$  &$-$0.07 
      & 5.744$\times 10^{-1}$           & 7.229$\times 10^{-1}$  \\ \hline
  150 & 2.799$\times 10^{-1}$  & $-$0.11& 2.800$\times 10^{-1}$  &$-$0.08 
      & 2.803$\times 10^{-1}$           & 3.588$\times 10^{-1}$ \\ \hline
  170 & 3.567$\times 10^{-2}$  & $-$0.13& 3.568$\times 10^{-2}$  &$-$0.11 
      & 3.573$\times 10^{-2}$           & 4.616$\times 10^{-2}$ \\ 
         \hline \hline
    & \multicolumn {6}{c|} {$\sigma_{mixed}$} \\ \hline
   10 & 2.470$\times 10^1$     &$-$0.20 & 2.475$\times 10^1$    &  +0.03 
      & 2.476$\times 10^1$              & 3.010$\times 10^1$    \\ \hline
   30 & 2.495                  &$-$0.16 & 2.500                 &  +0.00
      & 2.500                           & 3.088                 \\ \hline
   50 & 2.087                  &  +0.28 & 2.087                 &  +0.28
      & 2.080                           & 2.532                 \\ \hline
   70 & 2.545                  &  +0.10 & 2.546                 &  +0.13
      & 2.542                           & 3.134                 \\ \hline
   90 & 1.652                  &$-$0.05 & 1.653                 &  +0.00 
      & 1.653                           & 2.066                 \\ \hline
  110 & 7.697$\times 10^{-1}$  &$-$0.20 & 7.704$\times 10^{-1}$ &$-$0.12 
      & 7.716$\times 10^{-1}$           & 9.685$\times 10^{-1}$ \\ \hline
  130 & 5.572$\times 10^{-1}$  &$-$0.17 & 5.576$\times 10^{-1}$ &$-$0.11 
      & 5.584$\times 10^{-1}$           & 6.958$\times 10^{-1}$ \\ \hline
  150 & 8.648$\times 10^{-1}$  &$-$0.16 & 8.652$\times 10^{-1}$ &$-$0.13 
      & 8.666$\times 10^{-1}$           & 1.097                 \\ \hline
  170 & 1.204                  &$-$0.19 & 1.204                 &$-$0.19 
      & 1.207                           & 1.548                 \\ \hline
\end{tabular}
}\par
\medskip
{\footnotesize
\noindent{\bf Table B3}. Same as table B1, but for $\sqrt s = 200$ GeV.}

\newpage

\section*{Appendix C. The One-Loop Cross Sections in the High-Energy Approximation}
\setcounter{section}{3}
\setcounter{equation}{0}
     In this appendix, we give the formulae for cross sections in the Born-form
high-energy approximation and derive the linear relationship (\ref{3.11}) to
(\ref{3.13}) between the Born-form invariant amplitudes $\hat S_I^{(-)}(s,t)$ 
and $\hat S_Q^{(\pm)}(s,t)$ on the one hand and the cross-section coefficients
$C_{-,T}(s,t)$ and $C_{\pm,L}(s,t)$ from ref. \cite{BDDMS} on the other hand.
\par
The differential cross section for  given helicities $\sigma$, 
$\lambda$ and $\bar\lambda$ is given by
\bq
   {{d\sigma(\sigma,\lambda,\bar\lambda)}\over {d\cos\theta}}
   = {{\pi\beta}\over {2s^3}}\alpha^2
   \vert {\cal H}(\sigma,\lambda,\bar\lambda)\vert^2,
\label{C.1}
\eq
where the helicity amplitudes in the Born-form approximation become
\bq
    {\cal H}(\sigma,\lambda,\bar\lambda) 
    = S_I^{(\sigma)}(s,t) {\cal M}_I(\sigma,\lambda,\bar\lambda)
                            \delta_{\sigma,-} 
      + S_Q^{(\sigma)}(s,t) {\cal M}_Q(\sigma,\lambda,\bar\lambda).
\label{C.2}
\eq
For longitudinal $W$-bosons, with our choice of basic matrix elements
${\cal M}_i$ given in (\ref{A.1}) to (\ref{A.11}), the cross section depends
on $S_I^{(\pm)}(s,t)$ and $S_Q^{(\pm)}(s,t)$ only, i.e. relation (\ref{C.2}) 
becomes exact for $\sigma=- 1$, $\lambda=\bar\lambda=0$ rather than being 
an approximation.
Upon factorization into Born cross section and correction term, for
longitudinal $W$ polarization we have
\bq
  {{d\sigma}\over{d\Omega}}\ \bigg |_{\sigma, L}=
  {{d\sigma}\over{d\Omega}}\ \bigg |_{\sigma, L}^{Born}\ \cdot
      \left|\ 1 + 
        {{ {1\over{2s_W^2}}{\cal M}_I(\sigma,0,0) \delta\hat S_I^{(\sigma)}
          +{\cal M}_Q(\sigma,0,0) \delta\hat S_Q^{(\sigma)}  } \over
         { {1\over{2s_W^2}}{\cal M}_I(\sigma,0,0) \delta_{\sigma,-}
          +{\cal M}_Q(\sigma,0,0) } }\ \right| ^{\ 2},  
\label{C.3}
\eq
where $\delta \hat S_I^{(\sigma)}(s,t)$ and $\delta \hat S^{(\sigma)}_Q$ 
denote the deviations from their respective Born values:
\bqa
   \delta \hat S^{(-)}_I  &=&    \hat S^{(-)}_i -1, \nonumber\\
   \delta \hat S^{(+)}_I  &\equiv & {1\over {2s_W^2}} S^{(+)}_I,\\
   \delta S^\pm_Q         &=&   \hat S^{(\pm)}_Q -1.\nonumber
\label{C.4}
\eqa
We stress that (\ref{C.3}) does not yet involve any approximation.
Expanding the coefficients ${\cal M}_I$ and ${\cal M}_Q$ in powers of
$s$, the  cross section (\ref{C.3}) becomes
\bqa
  {{d\sigma}\over{d\Omega}}\ \bigg |_{-, L}&=&
     {{d\sigma}\over{d\Omega}}\ \bigg |_{-, L}^{Born} \cdot
     \left|\  1+ (2c_W^2-1)(1 + O({1\over s})) \delta \hat S_I^{(-)} 
      +  2s_W^2   (1 + O({1\over s})) \delta \hat S_Q^{(-)}\ \right|^2, 
     \label{C.5}\\ 
  {{d\sigma}\over{d\Omega}}\ \bigg |_{+, L}&=&
     {{d\sigma}\over{d\Omega}}\ \bigg |_{+, L}^{Born} \cdot
     \left|\  1+{{2c_W^2-1}\over{2s_W^2}}(1 + O({1\over s})) \delta \hat S_I^{(+)}
        + \delta \hat S_Q^{(+)}\ \right|^2. \label{C.6}
\eqa 
\par
A high-energy expansion for the production cross section at the one-loop
order was derived in ref. \cite{BDDMS}. It is given by
\bqa
  {{d\sigma}\over{d\Omega}}\ \bigg |_{-, L}&=&
     {{d\sigma}\over{d\Omega}}\ \bigg |_{-, L}^{Born} \cdot
     [1+ C_{-,L}^B + C_{-,L}^F ], \label{C.7}\\ 
  {{d\sigma}\over{d\Omega}}\ \bigg |_{+, L}&=&
     {{d\sigma}\over{d\Omega}}\ \bigg |_{+, L}^{Born} \cdot
     [1+ C_{+,L}^B + C_{+,L}^F ]. \label{C.8}
\eqa
 
Comparison of (\ref{C.5}), (\ref{C.6}) with (\ref{C.7}), (\ref{C.8})
yields relations (\ref{3.11}) and (\ref{3.12}).
\par
In the case of transverse polarization for the $W$-bosons, in 
contradistinction with (\ref{C.3}), the cross section also contains
contributions from invariant amplitudes different from
$S_I^{(\pm)}(s,t)$ and
$S_Q^{(\pm)}(s,t)$.  In this case, (\ref{C.2}) 
is indeed an approximation, even for left-handed electrons, $\sigma=-1$.
>From the numerical one-loop evaluation of appendix B we know that 
it is an excellent one, i.e. all leading terms are taken into account
by this high-energy Born-form approximation.
Moreover, according to table A1, equal 
transverse helicities are suppressed in contrast with opposite ones for
which ${\cal M}_Q$ is suppressed.  The one-loop corrected cross section
at high energies for transverse polarization of the $W$ bosons
depends on $\hat S_I^{(-)}(s,t)$ only:
\bq 
  {{d\sigma}\over{d\Omega}}\ \bigg |_{-, T}=
  {{d\sigma}\over{d\Omega}}\ \bigg |_{-, T}^{Born} \cdot 
          \vert 1 + \delta\hat S_I^{(-)}\vert^2.
\label{C.9}
\eq
Comparison with ref. \cite{BDDMS},
\bq
  {{d\sigma}\over{d\Omega}}\ \bigg |_{-, T}=
     {{d\sigma}\over{d\Omega}}\ \bigg |_{-, T}^{Born} \cdot
     [1+ C_{-,T}^B + C_{-,T}^F ], \label{C.10} 
\eq 
immediately implies relation (\ref{3.13}).
\par
In summary, with our choice of basic matrix elements, the high-energy
one-loop correction coefficients to cross sections, originally
introduced without reference to a Born-form approximation, become
identical to linear combinations of Born-form invariant amplitudes.

\vfill\eject

\vfill\eject

\end{document}